\documentclass[submission,Phys]{SciPost}
\usepackage{graphicx,bm}
\usepackage{amsmath}
\usepackage{amsfonts}
\usepackage{amssymb}
\usepackage{braket}
\usepackage{dsfont}
\usepackage{bm}
\usepackage{eucal}
\usepackage{dsfont}
\usepackage{mathtools}
\usepackage{comment}

\usepackage{bbm}

\usepackage{color}

\usepackage{xcolor}

\usepackage{tikz}
\usetikzlibrary{calc}
\usetikzlibrary{decorations.markings}
\usetikzlibrary{decorations.pathmorphing, arrows}
\usetikzlibrary{arrows.meta}
\usepackage{pgfplots}

\newcommand{\be}{\begin{equation}}
\newcommand{\ee}{\end{equation}}
\newcommand{\bea}{\begin{eqnarray}}
\newcommand{\eea}{\end{eqnarray}}

\def\nn{\nonumber\\}

\def\fr#1{(\ref{#1})}

\def\nn{\nonumber\\}

\def\fr#1{(\ref{#1})}

\def\abr{\alpha_{\rm BR}}

\def\txi{\widetilde{\xi}}

\def\sfix#1{\texorpdfstring{#1}{Lg}}
\def\wta{{\alpha}}
\graphicspath{{figs/}}

\newcommand{\paul}[1]{{\color{blue}\ifmmode\text{\footnotesize(PF) #1}\else\footnotesize{(PF) #1}\fi}}
\newcommand{\fab}[1]{{\color{red}\ifmmode\text{\footnotesize(FE) #1}\else\footnotesize{(FE) #1}\fi}}
\newcommand{\eric}[1]{{\color{green!40!black}\ifmmode\text{\footnotesize(EV) #1}\else\footnotesize{(EV) #1}\fi}}
\pgfplotsset{compat=1.18}
\begin{document}
\begin{center}
    \Large{\bf Strong zero modes in integrable spin-$S$ chains}
\end{center}
	\date{\today}
	\begin{center}
	F.H.L.\ Essler$^1$, P.\ Fendley$^{1,2}$, and E.\ Vernier$^3$, 
\end{center}	
\begin{center}
{\bf 1}
The Rudolf Peierls Centre for Theoretical Physics, University of Oxford, OX1 3PU, UK\\
{\bf 2} All Souls College, University of Oxford, UK\\
{\bf 3}  Université Paris Cité and Sorbonne Université, CNRS, Laboratoire de
  Probabilités, Statistique et Modélisation, F-75013 Paris, France
\end{center}
\section*{Abstract}
{\bf 
We derive exact strong zero mode (ESZM) operators for integrable spin-$S$ chains with open boundary conditions and a boundary field. Their locality properties are generally weaker than in the previously known cases, but they still imply infinite coherence times in the vicinity of the edges. We explain how such integrable chains possess multiple ground states describing a first-order quantum phase transition, and that the odd number of such states for integer $S$ makes the weaker locality properties necessary.  We make contact with more traditional approaches by showing how the ESZM for $S$\,=\,$\tfrac12$ acts on energy eigenstates given by solutions of the Bethe equations.}

\date{\today}
\tableofcontents

\section{Introduction}

Higher-spin generalizations of the Heisenberg and XXZ Hamiltonians play an important role in the physics of spin chains and the more general theory of phase transitions in two spacetime dimensions. Their phase diagrams display a multitude of interesting phenomena unseen in spin-$\tfrac12$ chains. 
Integrable spin chains provide one of the few useful tools for analytically understanding such physics when the couplings are strong. Early examples of higher-spin integrable chains were constructed in Refs \cite{zamolodchikov1980model,kulish1981yang,kulish1981generalized}, with some physical properties computed exactly in \cite{takhtajan1982picture,babujian1982exact,babujian1983exact,sogo1984ground,kirillov1986exact,kirillov1987exact,kirillov1987exactII,frahm1990integrable,suzuki2004excited,deguchi2010correlation}.
Integrable open boundary conditions were identified in Refs \cite{mezincescu1990bethe,de1994exact}.
While integrable higher-spin chains are hardly generic, they prove exceptionally useful in finding the structure of generic phase diagrams. Because of their increased symmetry, they often allow one to locate phase transitions exactly. For example, integrable chains describe several transition points in the $SU(2)$-invariant bilinear-biquadratic spin-1 chain \cite{affleck1987critical,Lauchli2006}.  
As such, they provide explicit lattice realizations of various interesting conformal field theories, for example for the $SU(2)_k$ Wess-Zumino-Witten universality classes. They also naturally describe first-order phase transitions. 

One particularly interesting phenomenon is the appearance of non-trivial edge modes. For example, they are a characteristic property of systems with topological order, with or without symmetry protection \cite{kitaev2001unpaired,Gu2009,Pollmann2012}. Remarkably, analogous edge modes remain at arbitrary energy densities in some systems \cite{fendley2012parafermionic,alicea2016topological,fendley2016strong,moran2017parafermionic,vasiloiu2019strong,kemp2020symmetry}, sometimes exactly, sometimes approximately. Such a \emph{strong zero mode} (SZM) operator maps all states in one symmetry sector to another, and (almost) commutes with the Hamiltonian. 
More precisely, a strong zero mode $\overline{\Psi}$ is an operator localized near one of the edges of an open system. It commutes with the Hamiltonian up to corrections exponentially small in the system size $L$, i.e.\ \cite{alicea2016topological} 
\begin{align}
    \big\| [H, \overline{\Psi}] \big\|={\cal O}(e^{-\alpha L})\qquad \hbox{for }\ L \rightarrow\infty\ .
\label{expsmall}
\end{align}
These corrections do not occur in all systems with an SZM, but when they do, they arise from the exponentially small tail of $\overline{\Psi}$ at the far edge. The SZM in all previous examples obeys $\overline{\Psi}^n \propto \mathds{1}$ as $L\to\infty$ for some integer $n > 1$. A consequence of the latter condition is that an SZM does not commute with some global symmetry generator of the theory, as it maps between different symmetry sectors. The former condition then shows that the energy splitting between the corresponding states must be exponentially small. 

Interesting physical consequences of this mapping include long-lived coherence of spins near the edge \cite{kemp2017long,else2017prethermal}.  Robust long-lived edge modes have recently been observed utilising superconducting processors \cite{SZMexperiment}.
Closely related phenomena occur in periodically driven systems, see e.g.\  \cite{Sen13,Chandran14,Bahri15,iadecola2015stroboscopic,Khemani16,sreejith2016parafermion,Yao17,Potter18,mukherjee2024emergent,vernier2024strong}. Interestingly, in both contexts these edge modes at finite energy densities have a degree of robustness under perturbations \cite{kemp2017long,yates2019almost,yates2020lifetime}. SZMs were found to also occur in stochastic processes \cite{klobas2023stochastic} and interfaces between different phases \cite{olund2023boundary}.

As was pointed out in Ref.~\cite{fendley2016strong}, it is possible to turn a SZM into an exact symmetry operator in a finite volume by changing the boundary conditions. For example, for an SZM localized around one edge, including a boundary field at the other edge allows the SZM to be modified into an {\em exact strong zero mode} (ESZM) operator $\Psi$ that commutes with $H$. If this modification breaks the discrete symmetry and removes the distinct sectors, degeneracies then no longer result. The ESZM therefore generates a very unusual symmetry. As we will explain in section~\ref{sec:auto}, ESZM and SZM operators have different effects on the behaviour of physical observables like boundary autocorrelation functions.

The goal of this paper is to understand better how such operators probe the physics of higher-spin integrable chains. We show that while operators similar to SZMs and ESZMs exist in higher-spin Heisenberg-like models with $U(1)$ symmetry, the situation is rather more complicated. In order to make the problem analytically tractable, we focus on the integrable spin-$S$ XXZ chain  \cite{zamolodchikov1980model,kulish1981yang,kulish1981generalized} with a family of integrable open boundary conditions. We find that the edge autocorrelators can exhibit long-term coherence in the same fashion as the spin-$\tfrac12$ case. Moreover, this behaviour is consistent with the presence of SZMs and ESZMs. However, by analysing the ground states, we show that the SZM for the spin-1 case must be qualitatively different from that for the spin-$\tfrac12$ case. We also provide evidence that SZMs for the spin-$\tfrac32$ case at least in some ways resemble more the spin-$\tfrac12$ analogs, suggesting there may be a meaningful distinction between SZMs for integer and half-integer integrable spin chains.

One important characteristic of the integrable spin-$S$ chain in its gapped antiferromagnetic phase is that it lies along the locus of a first-order quantum phase transition, as we explain in section \ref{sec:physical}. In particular, we characterize their degenerate ground states, and show how they can be understood as the transition between ordered phases (and for integer spin) disordered phases. In section~\ref{sec:auto} we present numerical results for infinite-temperature autocorrelation functions in the open integrable spin-$\tfrac12$ and spin-1 chains. We find anomalously long (conceivably infinite) coherence times in the vicinity of the boundary, reminiscent of the SZM physics. By including integrability-breaking perturbations we show that these exhibit a degree of robustness. In section \ref{sec:ESZM}, we construct ESZM operators from the family of commuting transfer matrices arising from integrability, and explain how they give rise to the long coherence times. In section~\ref{sec:BBS} we relate the ESZMs to {boundary bound state} solutions of the Bethe equations, which have been analyzed previously in some detail in the literature. Finally, we present our conclusions in section~\ref{sec:conclusions}.

\section{The physics of integrable higher-spin chains}
\label{sec:physical}

In this section we introduce integrable higher-spin generalizations of the XXZ spin chain. We focus on the physics in their gapped regions, where the effect of the ESZM is particularly apparent. Along a line of couplings in the antiferromagnetic gapped region, the integrable spin-$S$ chain exhibits $2S+1$ degenerate ground states with an elegant structure. These ground states are not related by any obvious symmetry, and it is thus natural to expect that this line describes phase coexistence. We show that indeed such structure is nicely understood as describing a first-order quantum phase transition in a larger (non-integrable) phase diagram.  

The spin-$S$ chains we discuss are comprised of 
$L$ 2$S$+1-state quantum systems, i.e.\ the Hilbert space is 
$\big(\mathbb{C}^{2S+1}\big)^{\otimes L}$. We consider only nearest-neighbor interactions built from operators $h_{j,j+1}$ acting non-trivially only on sites $j$ and $j+1$. Including boundary ``fields" $b^{(S)}_1$ and $b^{(S)}_{L}$ acting at the ends of the open chain, the open and periodic Hamiltonians can be written in the form
\begin{align}
    H^{(S)}_{\rm open}=b^{(S)}_1+b^{(S)}_{L}+\sum_{j=1}^{L-1} h^{(S)}_{j,j+1}\ ,\qquad
    H^{(S)}_{\rm per}=\sum_{j=1}^{L} h^{(S)}_{j,j+1}\ ,
\label{Hopenper}
\end{align}
where indices are interpreted mod $L$ in the latter. We say the open chain with vanishing boundary fields has ``free" boundary conditions.

For simplicity, we require a $U(1)$ symmetry in the higher-spin chains, and so describe integrable generalizations of the XXZ chain. We expect that our results generalize to the integrable higher-spin XYZ chain. The integrable spin-1 Hamiltonian was originally derived in \cite{zamolodchikov1980model}, and a systematic method for deriving all the integrable higher-spin Hamiltonians is based on the fusion of transfer matrices, some aspects of which are summarized in Appendix~\ref{app:Tsystem}.   Even with the $U(1)$ symmetry, the explicit expression for the Hamiltonian for arbitrary spin is rather nasty, and so we give it in the Appendix in equation  \fr{H_S}. These Hamiltonians depend on a single parameter $\eta$, chosen so that $\eta$\,=\,0 when the symmetry is enhanced to an $SU(2)$ antiferromagentic quantum critical point. 
The quantum-critical case has easy-plane anisotropy, which corresponds in our conventions to imaginary $\eta$. The gapped antiferromagnetic region of interest here corresponds to real $\eta>0$. For any $S$, a Kosterlitz-Thouless transition occurs at $\eta=0$.

\subsection{Spin \sfix{$\tfrac12$}}

We start by quickly reviewing results for the venerable spin-$\tfrac12$ XXZ chain. Defining $\sigma^a_j$ to be a Pauli matrix acting non-trivially on the $j$th site,  the Hamiltonian is built from
\begin{align}
h^{(\frac12)}_{j,j+1} = \sigma^x_j\sigma^x_{j+1}+\sigma^y_j\sigma^y_{j+1}
+\cosh(\eta)\big(\sigma^z_j\sigma^z_{j+1}-1\big)\ .
\label{hhalf}
\end{align}
For periodic or free boundary conditions, the Hamiltonians from \eqref{Hopenper} have a $U(1)$ symmetry generated by $\sum_j\sigma^z_j$, along with a $\mathbb{Z}_2$ spin-flip symmetry generated by $R=\prod_j \sigma^x_j$.
The symmetry is enhanced to $SU(2)$ at the KT transition point $\eta=0$. For real $\eta>0$ the interactions are strongly antiferromagnetic, and classic results (see e.g.\ \cite{Baxter1982}) show that the ground states exhibit N\'eel order. The order parameter obeys
\be
m_{\rm st}^2=\lim_{\ell\to\infty}\lim_{L\to\infty}\big\langle{\rm GS}\big|(-1)^\ell\sigma^z_j\sigma^z_{j+\ell}\big|{\rm GS}\big\rangle\neq 0.
\label{spinhalforder}
\ee
The spin-flip symmetry is therefore spontaneously broken when $L\to \infty$. The open chain with $L$ sites has two ground states whose energies differ by a term exponentially small in $L$, and are separated by a finite excitation gap from all other energy eigenstates. 

A strong zero mode occurs in the spin-$\tfrac12$ XXZ chain with free boundary conditions \cite{fendley2016strong} and $\eta$ real and positive. A nice expression for it terms of a matrix product operator (MPO) can be found in \cite{FGVV}, and is a special case of our formula \eqref{eq:TM} below. As $\Delta=\cosh\eta\to\infty$, the SZM localized at the left edge reduces simply to $\overline{\Psi}^{(\frac12)} =\sigma^z_1 + \mathcal{O}(1/\Delta).$  In this limit, the ground states are the two N\'eel states $\ket{+-+-\dots}$ and $\ket{-+-+\dots}$, and $\sigma^z_1$ indeed maps between the two combinations with eigenvalues $\pm 1$ of $R$. 

One then can obtain a series expansion for the SZM by splitting $H_{\rm free}$\,=\,$H_{xy}+\Delta H_z$, and then iterating order by order in $1/\Delta$.  Namely, the zeroth order term $\Psi_0\equiv\sigma^z_1$ does not commute with $H_{xy}$, but there exists an operator $\Psi_1$ such that $[H_{xy},\Psi_0]=-\Delta[H_z,\Psi_1]$. Considering then $\Psi_0+\Psi_1$ and iterating the procedure leads to a series expansion of the SZM \cite{fendley2016strong}:
\begin{align}
\overline{\Psi}^{(\frac{1}{2})}=&\sum_{s=0}^{\lfloor L/2\rfloor}\sum_{0<a_1<\dots<a_{2s}<b\leq L}\Delta^{-2(b-1)}\sigma^z_b\prod_{r=1}^s\psi\big(a_{2s-1},\,a_{2s}\big) 
\equiv \sum_{s=0}^{\lfloor L/2\rfloor} \Psi_s
\ ,
\cr
&\hbox{where }\ \psi(a,\,a')=\Delta^{a'-a}(1-\Delta^2)\left[\sigma^x_{a}\sigma^x_{a'}+\sigma^y_{a}\sigma^y_{a'}\right]\ .
\label{psihalf}
\end{align}
One needs to truncate the series once the terms stretch across the entire system. The commutator $[H_{xy},\,\Psi_{\lfloor L/2\rfloor}]$ does not vanish, but one can check that it is exponentially small in $L$. As manifest from \eqref{psihalf}, this SZM anticommutes with the spin-flip symmetry $R$. Less obviously, it squares to a constant, again up to exponentially small corrections. It thus indeed satisfies the conditions outlined in the introduction, and for $\Delta>1$ maps not only between the two ground states, but between all the eigenstates of the Hamiltonian (up to the exponentially small corrections). The entire spectrum of the $H^{(\frac12)}_{\rm free}$  with $\Delta>1$ thus forms (almost) doubly degenerate pairs with opposite eigenvalues of $R$. 


Including a boundary magnetic field at the ``far" end yields an ESZM \cite{fendley2016strong}. Namely, we define $H(h_L)$ by taking $b_L^{(\frac12)}=h_L \sigma^z_L$ in \eqref{Hopenper} while still keeping $b_1^{(\frac12)}=0$. We then have
\be
\big[H_{xy},\,\Psi_{\lfloor L/2\rfloor}\big] = -\Delta\big[H_z,B_{\rm far}\big],
\ee
where we define
\be
B_{\rm far}\equiv \Delta^{1-2L}\sum_{s=0}^{\lfloor L/2\rfloor}\sum_{0<a_1<\dots<a_{2s}\leq L}\;\prod_{r=1}^s\psi\big(a_{2s-1},\,a_{2s}\big)\ .
\label{Bdef}
\ee
The SZM then can be modified into an ESZM via
\be
\Psi^{(\frac12)}
\equiv\overline{\Psi}^{(\frac12)}+\frac{1}{h_L}B_{\rm far}\qquad\implies\qquad \Big[H^{(\frac12)}(h_L),\Psi\Big]=0\ .
\label{ESZMvsSZM}
\ee
Only the modification depends on $h_L$. A simple derivation of this ESZM follows from the MPO form, as we describe for general $S$ below. In the limit of vanishing boundary field $h_L\to 0$ the ESZM becomes ill-defined. 

In fact, given the ESZM we can prove the condition \eqref{expsmall} for $\overline{\Psi}$. 
We start by noting that
\be
\big[H(0),\overline{\Psi}^{(\frac12)}\big]=\big[B_{\rm far},\sigma^z_L\big]\ .
\ee
The definition \eqref{Bdef} requires that the only contributions to the commutator on the right-hand side are terms with $a_{2s}=L$. These terms necessarily have coefficients that at largest are order $\Delta^{L-1}$, yielding
\be
\big\Vert[H(0),\overline{\Psi}]\big\Vert={\cal O}(\Delta^{- L})\ .
\label{almostcommute}
\ee

It is illuminating to rewrite the model in terms of fermions.
(Almost) degenerate ground states in open fermionic chains are a characteristic of topological order. Both here and in the transverse-field Ising chain, a Jordan-Wigner transformation converts a spin chain into a fermion one, namely
\begin{align}
  \sigma^x_j=1-2c^\dagger_jc_j\ ,\qquad 
  \sigma^z_j=\prod_{\ell<j}(1-2c^\dagger_\ell c_\ell)\ (c_j+c_j^\dagger)\ .
\label{JWT}
\end{align} 
The non-locality of the map makes the antiferromagnetic order parameter from \eqref{spinhalforder} become non-local as well: 
\be
(-1)^\ell\Big\langle{\rm GS}\Big|(c_j+c^\dagger_j)\prod_{j\leq m<j+\ell}(1-2c^\dagger_m
c_m)\ (c_{j+\ell}+c_{j+\ell}^\dagger)\Big|{\rm GS}\Big\rangle\neq 0.
\ee
This non-local order in the fermion chain is known as topological order \cite{kitaev2001unpaired}.

In the limit $\eta\to\infty$, the Ising and spin-$\tfrac12$ XXZ chains become identical, and the strong zero mode is simply the edge fermion. Indeed, defining Majorana fermions via
\be
a_{2n}=-i(c_n^\dagger-c_n)\ ,\quad a_{2n-1}=c_n^\dagger+c_n\ ,
\ee
the leading term in the open Hamiltonian as $\eta\to\infty$ becomes
\be
H_0=2i\cosh(\eta)\sum_{n=1}^{L-1}a_{2n}a_{2n+1}\ .
\ee
The edge Majorana fermions then commute with this Hamiltonian
\be
\big[H_0,a_1\big]=0=\big[H_0,a_{L}\big].
\label{Majorana}
\ee
The strong zero modes from \cite{kitaev2001unpaired,fendley2016strong} indeed reduce to $a_1=\sigma^z_1$ and $a_L=\sigma^z_L$ in this limit.

\subsection{Spin 1}
\label{sec:spin1intro}


The integrable spin-1 XXZ chain was introduced by Zamolodchikov and Fateev in \cite{zamolodchikov1980model}. The Hamiltonian is written in terms of the usual spin-1 generators, which in the $s^z$-diagonal basis are
\be
s^x=\frac{1}{\sqrt{2}}\left(\begin{array}{c c c}0&1&0\\1&0&1\\0&1&0\end{array}\right),\quad s^y=\frac{1}{\sqrt{2}}\left(\begin{array}{c c c}0&-i&0\\i&0&-i\\0&i&0\end{array}\right),\quad s^z=\left(\begin{array}{c c c}1&0&0\\0&0&0\\0&0&-1\end{array}\right)\ .
\label{spin_op}
\ee
 The Hamiltonian densities are then
\begin{align}
h^{(1)}_{j,j+1} = \sum_{b} J_b \Big[s^{b}_js^{b}_{j+1} - \big(\big(s^{b}_{j}\big)^2-1\big)\big(\big(s^{b}_{j+1}\big)^{2}-1\big)+1\Big]
-\sum_{b,c|b\ne c}\mathcal{A}_{bc}~s^{b}_js^{c}_js^{b}_{j+1}s^{c}_{j+1},
\label{hspin1}
\end{align}
where $b,c\in\{x,y,z\}$ and $\mathcal{A}_{bc}=\mathcal{A}_{cb}$ depend on $\eta$ as
\be
J_x=J_y=A_{xy} =1\,,\quad J_z = \cosh\left(2\eta\right),\quad  \mathcal{A}_{xz}=\mathcal{A}_{yz}=2\cosh\eta-1\ .
\label{coefficients}
\ee
As above, we study the easy-axis antiferromagnetic regime $\eta>0$. The $U(1)$ symmetry is generated by $\sum_j s^z_j$, which commutes with each of the $h^{(1)}_{j,j+1}$ individually. The $\mathbb{Z}_2$ symmetry sends each $s^z_j\to - s^z_j$ and $s^y_j\to - s^y_j$, while leaving $s^x_j$ invariant. For any $\eta$ and appropriate boundary conditions, this chain exhibits a $N=2$ dynamical lattice supersymmetry  \cite{hagendorf2013spin}, similar to what happens at $\eta=\pi/3$ in the spin-$\tfrac12$ chain  \cite{yang2004non}.

\begin{figure}[ht]
\centering
\begin{tikzpicture}
  \begin{axis}[
    axis lines=middle,
    samples=300,
    domain=-1.3:1.3,
    ymin=0, ymax=0.5,
    ylabel={$V_{\rm LG}$},
    xtick=\empty,
    ytick=\empty,
    enlargelimits=true,
    width=12cm,
    height=4cm,
    clip=false,
  ]
    \addplot[blue, thick, smooth, domain=-1.25:1.25] {x^2*(x^2-1)^2};
  \end{axis}
  \node at (1.75,-0.5) {$|{\rm GS3}\rangle$};
  \node at (5.25,-0.5) {$|{\rm GS1}\rangle$};
  \node at (8.75,-0.5) {$|{\rm GS2}\rangle$};
\end{tikzpicture}
\caption{\small Triple-well Landau-Ginzburg potential.}
\label{fig:triplewell}
\end{figure}
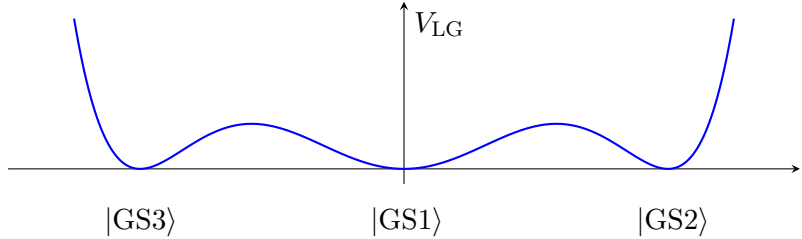

In this section we explain here how a triple-well potential provides a good effective picture for the low-energy states all along the gapped $\eta>0$ line of the integrable spin-1 XXZ chain with periodic or open boundary conditions. The three degenerate ground states, pictured as the three wells in fig.\ \ref{fig:triplewell}, are not related by any obvious symmetry: the $\mathbb{Z}_2$ spin-flip symmetry only reflects the picture, leaving the middle well invariant. Several other interesting models possess the same triple-well structure. The integrable hard-square lattice model and the ensuing quantum Hamiltonian are among them \cite{Baxter1982,Fendley2004}, with scaling limit described by the integrable $\Phi_{1,3}$ perturbation of the tricritical Ising conformal field theory \cite{Zamolodchikov1991}.   Finding such a triple-well structure does {\em not} however require imposing integrability in general. In the perturbed tricritical Ising example, imposing Kramers-Wannier duality is sufficient; the triple well occurs along the self-dual line. A lattice example of such can be found in \cite{OBrien2017}, where the duality manifestly maps between the three ground states.

One involved but powerful method for deriving this degeneracy in integrable models such as ours is to use the corner transfer matrix \cite{Andrews1984,Huse1984}. When a spin chain is the limit of an integrable two-dimensional classical model, this approach allows the computation of one-point functions in the presence of various boundary conditions. From these one can infer the ground-state spectrum. Our Hamiltonian is the limit of the integrable 19-vertex model, and one indeed obtains results that imply the triple-well structure \cite{Date1988}.  

A more direct way of understanding the physical properties of the integrable spin-1 model \eqref{hspin1} comes from considering the limits of very strong ($\eta\to\infty$) and very weak ($\eta\to 0^+$) anisotropy. We show next how the triple well arises in both limits. 

\subsubsection{Strong coupling analysis (\sfix{$\eta\to\infty$})}
In the regime $\eta\gg 1$ we expand the Hamiltonian in a power series in $e^{-\eta}$. The periodic Hamiltonian in this region can be expanded as 
$H^{(1)}_{\rm per}={e^{2\eta}}H_0+e^{\eta}H_1+H_2+\dots $, where the first two terms are
\begin{align}
H_0&=\tfrac12\sum_{j=1}^{L}\big[s^z_js^z_{j+1}- \big((s^{a}_{j})^2-1\big)\big((s^{a}_{j+1})^{2}-1\big)+1\big]\ ,\nn
H_1&=\sum_{j=1}^{L}\Big[\big(s^x_js^z_j\big)\big(s^x_{j+1}s^z_{j+1}\big)+\big(s^y_js^z_j\big)\big(s^y_{j+1}s^z_{j+1}\big)+{\rm h.c.}\Big].
\label{H1_SC}
\end{align}
The leading term $H_0$ has three degenerate ground states with $E=0$: 
\be
|{\rm GS},1\rangle=|0,0,\dots ,0\rangle\ ,\quad
|{\rm GS},2\rangle=|+,-,+,-,\dots\rangle\ ,\quad
|{\rm GS},3\rangle=|-,+,-,+,\dots\rangle\ .
\ee
Here we use notations where
\begin{align}
s^z_j|\tau\rangle_j=\tau|\tau\rangle_j\ ,\quad \tau\in\{0,\pm\}\ ,\quad
|\tau_1,\dots,\tau_L\rangle=\otimes_{j=1}^L|\tau_j\rangle_j.
\end{align}
No obvious symmetry mixes $|{\rm GS},1\rangle$ with the other two ground states. 

At finite $L$,  these ground states mix only at $L/2$th order in perturbation theory, so that the splitting at large $\eta$ is order $e^{-\eta L}$. One finds that the linear combinations giving eigenstates of the full Hamiltonian in this limit are
\begin{align}
|{\rm GS},\pm\rangle&=\frac{1}{\sqrt{2}}|{\rm GS},1\rangle\pm\frac{1}{2}\big(|{\rm GS},2\rangle+|{\rm GS},3\rangle\big)\ ,\nn
|{\rm GS},\pi\rangle&=\frac{1}{\sqrt{2}}\big(|{\rm GS},2\rangle-|{\rm GS},3\rangle\big)\ .
\end{align}
All three of these states have off-diagonal long-range N\'eel order,
as defined by the order parameter
\be
\mathcal{N}_a=\lim_{\ell\to\infty}\lim_{L\to\infty} \big\langle {\rm GS},a\big|(-1)^\ell s^z_js^z_{j+\ell}\big|{\rm GS},a\big\rangle\ .
\ee
At leading order, $\mathcal{N}_0=2\mathcal{N}_\pm$. 

Excited states are conveniently written in terms of domain walls. An interesting feature of the integrable higher-spin chain is that not all domain walls have the same energy. For $H_0$, the lowest-energy domain walls are between $|{\rm GS},1\rangle$ and either of other two ground states, with energy $\tfrac12 e^{2\eta}$ relative to the ground state.   Thus with periodic boundary conditions, the $L(L-2)$ lowest-lying eigenstates of $H_0$ are of the form
\begin{align}
&|0,\dots ,0,+,-,\dots,+,-,0\dots0\rangle\ ,\qquad
|0,\dots ,0,-,+,\dots,-,+,0\dots0\rangle\ .
\end{align}
They have $S^z=0$ and $\Delta E=e^{2\eta}$. The domain wall between $|{\rm GS},2\rangle$ and $|{\rm GS},3\rangle$ has energy $e^{2\eta}$, twice the energy of those involving $|{\rm GS},1\rangle$. Thus it is natural to think of the former as a pair of the latter. Such a low-energy theory can therefore be described by a triple-well Landau-Ginzburg effective potential from Fig.~\ref{fig:triplewell}.
The lowest-energy domain walls separate adjacent ground states in the picture. The lack of symmetry between $|{\rm GS},1\rangle$ and the other two is apparent.

Because no obvious symmetry guarantees this triple degeneracy, one might expect that this elegant low-energy theory is simply an artefact of the large-$\eta$ limit. However, 
second-order perturbation theory in $e^{-\eta}$ (which requires taking into account the next term $H_2$ in the expansion of $H^{(1)}$) does not remove the ground-state degeneracy. We explain below how such degeneracy persists all along the line with $\eta>0$, making the triple-well effective potential describe the low-energy theory throughout this region.

Before doing so, we note how boundary conditions can affect the low-energy structure dramatically. 
Including the boundary fields
\begin{align}
b^{(1)}_1=(s^z_1)^2,\qquad b^{(1)}_L=(s^z_L)^2
\end{align}
into the open chain from \eqref{Hopenper} gives an integrability-preserving example.
These boundary fields manifestly break the three-fold ground state degeneracy and make $|{\rm GS},1\rangle$ the unique ground state. 
The lowest-lying excited states have energy $E_1=e^{2\eta}$ and are $L^2/2$-fold degenerate. They are states with
\begin{itemize}
\item{} no domain walls: $|+-+-\dots\rangle$, $|-+-+\dots\rangle$,
\item{} one domain wall:
$|0\dots 0+-+-\dots\rangle$, $|0\dots 0-+-+\dots\rangle$, $|+-\dots+-0\dots 0\rangle$, $|-+\dots-+\,0\dots 0\rangle$,
\item{} two domain walls: $|0\dots 0+-\dots +-\,0\dots 0\rangle$.
\end{itemize}

\subsubsection{Field-theory limit of the weakly anisotropic spin-1 chain}
\label{app:RG}

The same low-energy picture emerges in the effective field-theory description valid at small $\eta$. In this limit, the low-energy sector of the integrable spin-1 XXZ chain \eqref{hspin1} is described by the supersymmetric sine-Gordon field theory  \cite{di1977classical,ahn1991complete,ahn2007finite}. Here we review this result. 

In the vicinity of the quantum-critical $SU(2)$-invariant point at $\eta=0$, the low-energy theory can be written in terms of three Ising field theories \cite{tsvelik1990field}. Describing Ising degrees of freedom by three sets of chiral Majorana fermion operators $L_a(x),\,R_a(x)$ with $x\propto j$, the Hamiltonian density is
\begin{align}
{\cal H}&=
\sum_{a=1}^2\Big[\tfrac12{iv_\parallel}\big(L_a\partial_xL_a-R_a\partial_xR_a\big)
-im_\parallel R_aL_a\Big]+g_\parallel J^3J^3\nn
&\quad+\frac{iv_\perp}{2}\big(L_3\partial_xL_3-R_3\partial_xR_3\big)
-im_\perp R_3L_3+g_\perp (J^1J^1+J^2J^2)+\dots\ ,
\label{Hfield}
\end{align}
where the $\dots$ includes only irrelevant operators, and 
\be
J^a=-\frac{i}{2}\epsilon^{abc}[L_bL_c+R_bR_c].
\ee
The relations between the original spin operators and Ising fields are 
\begin{gather}
S^\alpha_j\sim J^\alpha(x)+(-1)^{j} n^\alpha(x)\ ,\quad
n^x\sim \sigma^1\mu^2\mu^3\ ,\quad
n^y\sim \mu^1\sigma^2\mu^3\ ,\quad
n^z\sim \mu^1\mu^2\sigma^3\ ,
\label{oprelations} 
\end{gather}
where $\sigma^a$ and $\mu^a$ are Ising order and disorder fields. 
In writing \fr{Hfield} we have exploited the $U(1)$ symmetry generated by the current $J^3$. The full symmetry includes 
as well $\mathbb{Z}_2$ spin-flip symmetry along with translation invariance for periodic boundary conditions \cite{essler2004haldane}. 

A crucial consequence of the integrability of the spin chain is that only a single one of the relevant operators is allowed in the effective field theory. This simplification follows from computing the excitation gap in the integrable spin chain \cite{sogo1984ground}.  Tuning $\eta$, a quantum phase transition of occurs at the SU(2)-invariant point $\eta=0$, with vanishing gap for $\eta<0$ and an exponentially small gap for $\eta>0$. Such a transition is therefore of Kosterlitz-Thouless type, and the mass terms in the effective field theory must be fine-tuned to vanish in the integrable model:
\be
m_\perp=m_\parallel=0.
\ee
The only allowed interactions are governed by $g_\perp$ and $g_\parallel$. Bosonizing two of the Majorana fermions yields shows that \eqref{Hfield} with $m_\perp=m_\parallel=0$ is precisely the supersymmetric sine-Gordon field theory \cite{di1977classical,ahn1991complete}. The bosonic field $\Phi$ is defined via \cite{gogolin2004bosonization}
\bea
i[R_1L_1+R_2L_2]&\sim&\frac{1}{\pi\alpha}\cos\sqrt{4\pi}\Phi\ ,\qquad
i[L_1L_2+R_1R_2]\sim\frac{1}{\sqrt{\pi}}\partial_x\Phi\ ,\nn
L_1+iL_2&\sim&\frac{1}{\sqrt{\pi\alpha}}e^{-i\sqrt{4\pi}\varphi_L}\ ,\qquad
R_1+iR_2\sim\frac{1}{\sqrt{\pi\alpha}}e^{i\sqrt{4\pi}\varphi_R}\ ,
\eea
where $\alpha$ is a short-distance cutoff. Defining the dual field by
$
\Theta=\varphi_L-\varphi_R 
$
yields the Hamiltonian density
\be
{\cal H}=\frac{iv_\perp}{2}\left[L_3\partial_xL_3-R_3\partial_xR_3\right]
+\frac{v'_\parallel}{2}\left[\frac{1}{K}(\partial_x\Phi)^2+K(\partial_x\Theta)^2\right]
-\frac{2ig_\perp}{\pi\alpha}\cos\sqrt{4\pi}\Phi\ L_3R_3.
\label{HSSG}
\ee
Here the Luttinger parameter and renormalized velocity are respectively $K=(1+\frac{g_\parallel}{4\pi})^{-1/2}$ and $v'_\parallel=v_\parallel/K$. The renormalization group flow for $K$ and $g_\perp$ is indeed of Kosterlitz-Thouless form \cite{gogolin2004bosonization}, and exhibits a gapped strong coupling phase for $g_\parallel>0$.
The $g_\perp$-term in \fr{HSSG} must therefore be included in the effective Hamiltonian, yielding precisely the massive supersymmetric sine-Gordon field theory. 

The triple-well structure of this field theory is not immediately obvious in its Hamiltonian. Insight comes from making the simple mean-field approximation
\be
\cos\sqrt{4\pi}\Phi\,L_3R_3\longrightarrow
\big\langle\cos\sqrt{4\pi}\Phi\big\rangle\, L_3R_3+
\cos\sqrt{4\pi}\Phi\,\big\langle L_3R_3\big\rangle\ .
\ee
This decoupling is compatible with the $U(1)\rtimes \mathbb{Z}_2$ symmetry, and gives a mean-field Hamiltonian density ${\cal H}_{\rm MF}={\cal H}_{\rm I}+{\cal H}_{\rm SG}$, where
\begin{align}
{\cal H}_{\rm I}&=\frac{iv_\perp}{2}\left[L_3\partial_xL_3-R_3\partial_xR_3\right]-im R_3L_3\ ,\nn
{\cal H}_{\rm SG}&=
\frac{v'_\parallel}{2}\left[\frac{1}{K}(\partial_x\Phi)^2+K(\partial_x\Theta)^2\right]
+g\cos\sqrt{4\pi}\Phi\ .
\end{align}
Here the mass $m$ and the Sine-Gordon coupling are determined self-consistently as
\be
m=-\frac{2g_\perp}{\pi\alpha}\langle\cos\sqrt{4\pi}\Phi\rangle  ,\qquad g=\frac{2ig_\perp}{\pi\alpha}\langle R_3L_3\rangle\ .
\ee
The two decoupled Hamiltonians ${\cal H}_{\rm I}$ and ${\cal H}_{\rm SG}$ are those of the Ising and sine-Gordon field theories respectively.
We then must distinguish between two cases:
\begin{itemize} 
\item{} $m<0$. In this case the Ising field theory is in its ordered phase and
\[
\langle \Phi\rangle=\sqrt{\frac{\pi}{4}}\ ,\quad
\langle\sigma^3\rangle\neq 0\ .
\]
Given \eqref{oprelations} and the operator product
\[
\mu^1(x)\mu^2(x)\sim \sin\big(\sqrt{\pi}\Phi(x)\big)\ ,
\]
the mean-field theory predicts a non-vanishing staggered magnetization in z-direction $\langle n^z(x)\rangle\neq 0$.
There are two mean-field solutions corresponding to opposite signs of the staggered magnetization.
\item{} $m>0$. In this case the third Ising model is in its disordered phase and
\be
\langle\Phi\rangle=0\ ,\quad \langle\mu^3\rangle\neq 0.
\ee
The relations \fr{oprelations} between scaling fields and lattice operators imply that no magnetic order is present here.
\end{itemize}
The N\'eel and disordered ground states thus arise from different mean-field approximations, with the former for $m<0$, and the latter for $m>0$. It is thus plausible for all three ground states to be possible in the full theory. In this picture the domain walls discussed above separate regions with different expectation values $\langle\cos\sqrt{4\pi}\Phi\rangle$ and $\langle\sigma^3\rangle$. 

Since this model is integrable, exact scattering theory for the field theory provides a way to go beyond mean-field theory. Similarly to the analysis of the perturbed tricritical Ising model \cite{Reshetikhin1989,Zamolodchikov1991}, the field theory has three ground states not related by any obvious symmetry.  Moreover, low-energy excitations are described precisely by the same domain walls as we derived for the $\eta\to\infty$ limit \cite{ahn1991complete}.

\subsubsection{The \sfix{$\eta>0$} line as a first-order phase transition}

Multiple ground states unrelated by any obvious symmetry are characteristic of a first-order transition. We here show how the integrable spin-1 XXZ chain in its gapped antiferromagnetic region describes a first-order transition line between N\'eel ordered and disordered phases.

A simple Hamiltonian for $S=1$ displaying such behaviour is
\be
H^{(1)}(\eta,D)=H^{(1)} + D \sum_{j=1}^L (s^z_j)^2\ .
\label{HD}
\ee
Such a term obviously favors $|{\rm GS},1 \rangle$ for $D>0$ and the other two for $D<0$. Thus for large enough $\eta$, $H^{(1)}(\eta,D)$ with $D>0$ has a unique ground state not spontaneously breaking any symmetries. However, for $D<0$ the spin-flip symmetry is broken, yielding two ground states with N\'eel order. 


The $D<0$ N\'eel-ordered phase possesses a fermionic edge zero mode, just as the spin-$\tfrac12$ case discussed above.  This is easily seen in a fermionic representation valid for $D\to-\infty$. 
In this limit, the low-energy sector is obtained by restricting just to the spins with $s^z=\pm 1$. In this subspace spin-1 operators act as Pauli matrices, namely 
\begin{gather}
s^z_j\rightarrow\sigma^z_j\ ,\qquad s^{x,y}_j\rightarrow 0\ ,\qquad
s^x_js^y_j\rightarrow\frac{1}{2}(\sigma^y_j+i\sigma^z_j)\ ,\nn
(s^x_j)^2\rightarrow \frac{1}{2}(1+\sigma^x_j)\ ,\qquad (s^y_j)^2\rightarrow \frac{1}{2}(1-\sigma^x_j).
\end{gather}
The Hamiltonian in this subspace then reduces to that of the spin-$\tfrac12$ XXZ chain:
\be
H_{\rm eff}=-\frac{1}{2}\sum_j \left[\sigma^x_j\sigma^x_{j+1}+\sigma^y_j\sigma^y_{j+1}-\big(3+4\sinh^2\eta\big)\sigma^z_j\sigma^z_{j+1}\right]+{\rm const}.
\ee
The Jordan-Wigner transformation \fr{JWT} yields an edge mode as described above.

At large enough $\eta$, the transition between the N\'eel and disordered phases at $D=0$ is direct.
The field theory valid for $\eta\to 0^+$ also has a direct transition, as the now-present masses $m_\perp$ and $m_\parallel$ are relevant. Hence it is natural to expect that the integrable chain describes a line of first-order quantum phase transitions for all $\eta>0$. 

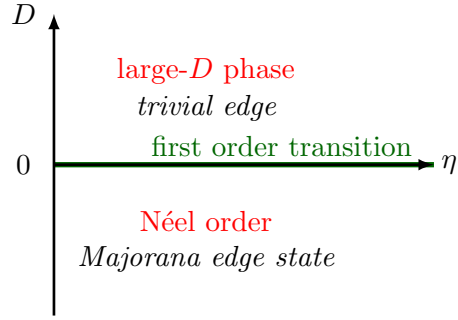
\begin{figure}[ht]
\centering
\begin{tikzpicture}
    \draw[line width=2,green!40!black] (0,0) -- (5,0);
    \node[green!40!black] at (3,0.25) {first order transition};

    \draw[line width=1,->,>=latex] (0,-2) -- (0,2);
    \draw[line width=1,->,>=latex] (0,0) -- (5,0);
    \node at (-0.4,0) {$0$};
    \node at (-0.4,2) {$D$};
     \node at (5.2,0) {$\eta$};
\node[red] at (2,1.25) {large-$D$ phase};
\node at (2,0.75) {\it trivial edge};

\node[red] at (2,-0.75) {Néel order};
\node at (2,-1.25) {\it Majorana edge state};

\end{tikzpicture}
\caption{\small Ground state phase diagram of the integrable $S=1$ chain perturbed by a single-ion anisotropy term \fr{HD}.}
\label{fig:GSPD}
 \end{figure}

To elaborate on the situation near $\eta=0$, we note that including the $D$-term in the field theory gives rise to a strongly relevant perturbation. The perturbation is generated by the $(n^z)^2$-term obtained by using the relations \fr{oprelations} between lattice operators and field, and is
\be
\delta {\cal H}=-i\lambda'\int dx \left[R_3L_3-R_1L_1-R_2L_2\right]\ .
\label{deltaH}
\ee
with $\lambda' \propto D$. The relative sign arises from the sign difference in the sub-leading terms in the $\mu\mu$ and $\sigma\sigma$ OPEs \cite{Ginsparg1988}.
The low-energy physics is then (approximately) described by three \emph{massive} Ising field theories, where the masses are renormalized by the marginal current-current interaction. The signs in \fr{deltaH} are such that
\begin{itemize}
\item{} For $\lambda'<0$ Ising models 1 and 2 are in their disordered phases $\langle\mu^{1,2}(x)\rangle  \neq 0$, while
Ising model 3 is in its ordered phase $\langle\sigma^3(x)\rangle\neq 0$. This implies that there is N\'eel order along the z-axis
\be
\langle n^z(x)\rangle=\langle\mu^1(x)\mu^2(x)\sigma^3(x)\rangle\neq 0\ .
\ee
All three Ising field theories are massive, resulting in a finite excitation gap. The fact that the third is in its ordered phase results in a Majorana edge state on the open chain.
\item{} For $\lambda'>0$, the situation is reversed: The third is in its disordered phase $\langle\mu^3(x)\rangle\neq 0$, while the other two 
are in their ordered phases $\langle\sigma^{1,2}(x)\rangle  \neq 0$.
The latter expectation values do not translate into magnetic order in terms of the original lattice spin operators given their relations \fr{oprelations} to the continuum fields, but instead indicate of a large-$D$ phase. 
As all three Ising models are massive there is a finite excitation gap.
\end{itemize}
Altogether we arrive at the ground state phase diagram shown in Fig.~\ref{fig:GSPD}.

\subsection{Spin \sfix{$\tfrac32$} and beyond}
\label{sec:spin32}

We expect that the integrable XXZ chain describes a first-order transition line for all $\eta>0$ for any $S>\tfrac12$. Corner transfer matrices \cite{Date1988} and exact scattering calculations indicate that the ground-state degeneracy becomes $2S+1$-fold \cite{Bernard1990}.

Some intuition into this result comes from probing the chains at small and large $\eta$. As with $S=1$, all relevant operators for $\eta\sim 0$ save the marginally relevant one are forbidden from the effective theory because of the $U(1)$ symmetry and the exponential smallness of the excitation gap \cite{sogo1984ground}. The effective field theory therefore is the $SU(2)_{2S}$ WZW model perturbed by a marginally relevant current-current interaction \cite{affleck1987critical}. In the integrable model all other relevant perturbations are fine-tuned to zero. The corresponding perturbed conformal field theory has indeed $2S+1$ degenerate vacua \cite{Bernard1990}.  There are a number of proximate gapped phases obtained by considering perturbations of this model by one of its relevant operators \cite{affleck1987critical}. 

Here we show that the integrable $S=\tfrac32$  chain in the large-$\eta$ limit indeed has four degenerate ground states. The leading term in \fr{H_S} for periodic boundary conditions for $S=\tfrac32$ and $L$ even is
\begin{align}
H_0^{(\frac32)}=\tfrac{1}{4}\sum_j\Big[&2P_{j,j+1}^{11}+2P_{j,j+1}^{44}
+P_{j,j+1}^{12}
+P_{j,j+1}^{21}+P_{j,j+1}^{34}+P_{j,j+1}^{43}\nn&
-P_{j,j+1}^{14} -P_{j,j+1}^{41} -P_{j,j+1}^{23} -P_{j,j+1}^{32}
 -3\Big],
\label{H032}
\end{align}
where we define basis states $\ket{1},\ket{2},\ket{3},\ket{4}$ on site $j$ such that
\be
s^z_j|a_j\rangle=\big(\tfrac52-a_j\big)|a_j\rangle\ ,\quad 
P_{j,j+1}^{ab} = \big|a_j\,b_{j+1}\big\rangle\big\langle a_j\,b_{j+1}\big|\ .
\ee
For periodic boundary conditions and even $L$, $H^{(\frac32)}_0$ has four degenerate ground states given by
\be
|{\rm GS},a\rangle=|a,5-a,a,5-a,\dots\rangle\ ,\quad a=1,2,3,4.
\ee
All these states exhibit antiferromagnetic off-diagonal long-range order in the sense that
\be
\lim_{\ell\to\infty}\lim_{L\to\infty}\frac{(-1)^\ell}{L-\ell}\sum_{j=1}^{L-\ell}\langle
s^z_js^z_{j+\ell}\rangle\neq 0.
\ee

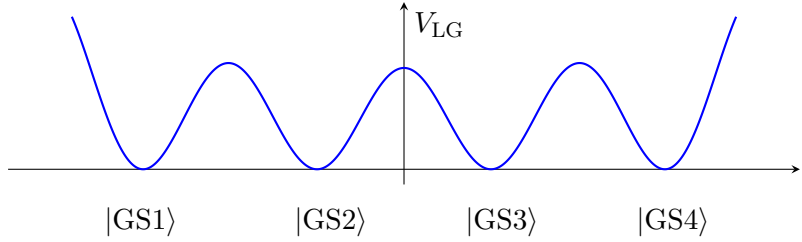
\begin{figure}[ht]
\begin{center}
\begin{tikzpicture}
  \begin{axis}[
    axis lines=middle,
    samples=300,
    domain=-3:3,
    ymin=0, ymax=1.5,
    ylabel={$V_{\rm LG}$},
    xtick=\empty,
    ytick=\empty,
    enlargelimits=true,
    width=12cm,
    height=4cm,
    clip=false,
  ]
    \addplot[blue, thick, smooth, domain=-3:3] {(1 + 0.0078125*x^4)*(cos(deg(2*x)))^2};
  \end{axis}
  \node at (1.75,-0.5) {$|{\rm GS1}\rangle$};
  \node at (4.25,-0.5) {$|{\rm GS2}\rangle$};
  \node at (6.5,-0.5) {$|{\rm GS3}\rangle$};
  \node at (8.75,-0.5) {$|{\rm GS4}\rangle$};
\end{tikzpicture}
\caption{\small Landau-Ginzburg potential describing the four ground states for $S=3/2$. Low-lying excitations consist of domain walls between between neighbouring minima.}
\label{fig:quadruplewell}
\end{center}
\end{figure}

The effective Landau-Ginzburg potential must have four degenerate wells. To find out how to arrange them, we find the lowest-energy excited states. They are easiest to write in terms of domain walls. As with the spin-1 case, not all domain walls have the same energy. The lowest-energy walls have energy 1 relative to the ground state, and correspond to nearest-neighbor pairs of states that do not appear in \eqref{H032}, namely
\be
(1,3),\ (2,2),\ (2,4),\ (3,1),\ (3,3),\ (4,2).
\ee
The simplest domain walls separate regions of adjacent ground states, so the effective potential must look as in figure \ref{fig:quadruplewell}.
Moreover, these domain walls are degenerate even when not symmetry related, i.e.\ the gap for the wall between ground states 2 and 3 is the same as that for the wall between 1 and 2 (and 3 and 4). Note as well that the energies of other nearest-neighbor pairs from \eqref{H032} also can be read off from the picture, e.g.\ a 11 pair has energy 3 relative to the ground state, the same as that that of a 3 domain-wall state.  
This picture extends to all $S$: the effective potential has $2S+1$ minima in a line, with degenerate domain walls interpolating between the adjacent ground states. No obvious symmetry causes the degeneracy.

\begin{figure}[ht]
\centering
\begin{tikzpicture}
    \draw[line width=2,green!40!black] (0,0) -- (5,0);
    \node[green!40!black] at (3,0.25) {first order transition};

    \draw[line width=1,->,>=latex] (0,-2) -- (0,2);
    \draw[line width=1,->,>=latex] (0,0) -- (5,0);
    \node at (-0.4,0) {$0$};
    \node at (-0.4,2) {$D$};
     \node at (5.2,0) {$\eta$};
\node[red] at (2,1.25) {Néel order I};
\node at (2,0.75) {\it Majorana edge state};

\node[red] at (2,-0.75) {Néel order II};
\node at (2,-1.25) {\it Majorana edge state};

\end{tikzpicture}
\caption{\small Ground state phase diagram of the model \fr{HD} for $S=3/2$.}
\label{fig:GSPD32}
 \end{figure}
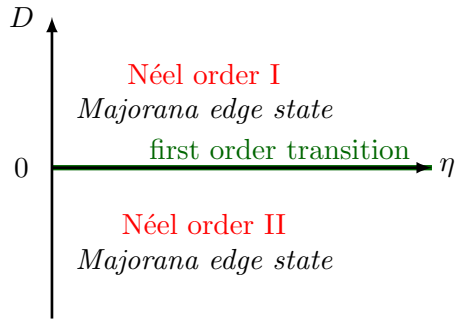

To see how the integrable model describes a first-order transition line, we include the single-ion anisotropy $D\sum_j(s_j^z)^2$ as in \fr{HD}. We find that the integrable line is bordered by two gapped, antiferromagnetically ordered phases. 
At positive $\eta$, non-zero $D$ removes the four-fold ground-state degeneracy, yielding the phases
\begin{itemize}
\item{} $D>0$: N\'eel-ordered phase with a doubly degenerate ground state
\be
|2,3,2,3\dots\rangle\ ,\
|3,2,3,2\dots\rangle\ ,
\ee
and a finite excitation gap. Similarly to $S$\,=\,1, the low-energy sector at $D\to\infty$ is obtained by projecting onto the $s^z_j=\pm \tfrac12$ states and the effective Hamiltonian takes the form of a spin-$\tfrac12$ XXZ chain
\begin{align}
H_{\rm eff}&=J\sum_j \left[\sigma^x_j\sigma^x_{j+1}+\sigma^y_j\sigma^y_{j+1}+\Delta\sigma^z_j\sigma^z_{j+1}\right]+{\rm const}\ ,\nn
J&=\frac{(\cosh\eta)^2}{1+2\cosh(2\eta)}\ ,\quad
\Delta=2 \cosh(\eta) + 3 \cosh(3\eta) +\frac{1}{2\cosh\eta}.
\end{align}
The open chain in this limit features a Majorana edge mode, as follows from the Jordan-Wigner transformation \fr{JWT} for the spin-$\tfrac12$ case.
\item{} $D<0$: N\'eel-ordered phase with two degenerate ground states
\be
|1,4,1,4\dots\rangle\ ,\
|4,1,4,1\dots\rangle\ ,
\ee
and a finite excitation gap. For open boundary conditions, the analogous large-$D$ calculation yields a Majorana edge mode as with $D>0$.
\end{itemize}
We thus arrive at the ground state phase diagram shown in Fig.~\ref{fig:GSPD32}.

Boundary conditions on an open chain can change this effective picture. One integrable set of boundary conditions is written at large $\eta$ as
\begin{align}
H^{(\frac32)}_B= \frac{e^{3\eta}}{4}[P_1^{2}+P_1^{3}+P_L^{2}+P_L^{3}]\ ,
\end{align}
where $P_j^a=|a_j\rangle\langle a_j|$.
The ground state for $D=0$ then is doubly degenerate, with only $|{\rm GS},1\rangle$ and $|{\rm GS},4\rangle$ surviving the perturbation. The lowest-lying excited states have excitation energy $\Delta E=e^\eta/2$ and come in three different types. The first type consists of the former ground states $|{\rm GS},2\rangle$ and  $|{\rm GS},3\rangle$ with no domain walls. The second type has one domain wall, e.g.\ $ |41\dots4132\dots 32\rangle$. The third type has two domain walls, e.g.\  $|41\dots4132\dots 3241\dots 41\rangle$.

\section{Edge coherence times}
\label{sec:auto}

Strong zero modes localized around the boundary have proved to be valuable probes of 
the physics of the edge region in spin-$\tfrac12$ chains, see e.g.\ \cite{fendley2016strong,kemp2017long, else2017prethermal,kemp2020symmetry}. They give rise to anomalous behaviour of observables in the boundary region, both in and out of equilibrium. A key characteristic of the edge-localized SZM in the spin-$\tfrac12$ XYZ chain is the pairing between eigenstates in different sectors, where the split between two energies is exponentially small in $L$. A long edge-spin coherence time is one of the consequences of such pairing. 

The presence of an odd number of degenerate ground states for the integrable integer-spin chain (see fig.\ \ref{fig:triplewell} for $S$\,=\,1) means that such a simple pairing scenario cannot apply here. Any analog of the SZM or ESZM must have more complicated behaviour.  Our explicit construction in sec.\ \ref{sec:ESZM} indeed finds spin 1 is qualitatively distinct from that of spin $\tfrac12$. For example, the edge-localization properties are weaker. However, these operators remain normalizable even as $L\to\infty$, suggesting that edge physics will exhibit special properties.

As the construction and analysis of the SZM and ESZM for spin 1 is rather technical, in this section we first gain some qualitative understanding of the edge physics. We use numerical computations to analyze various aspects of the edge physics of the integrable spin-$\tfrac12$ and spin-1 chains. We show that despite some differences between the two, many features are similar. A useful diagnostic for the presence of both EZSMs and SZMs are finite-temperature autocorrelation functions
\be
C^{\cal O}_\beta(t)=\frac{1}{Z(\beta)}{\rm Tr}\left[e^{-\beta H}{\cal O}(t){\cal O}(0)\right],
\label{autocorr}
\ee
where $Z(\beta)={\rm Tr}(e^{-\beta H})$ and ${\cal O}(0)$ are local operators that act non-trivially only very close to the left boundary. In the following we will only consider the particular case of infinite temperatures, where $Z(0)=(2S+1)^L$. In generic
spin chains such correlation functions decay quickly in time to asymptotic values that go to zero as the system size increases. 
For the $\mathds{Z}_2$-symmetric cases where SZMs appear, we address the following questions:
\begin{itemize}
\item{} Given that the integrable spin-$S$ chains exhibit a $U(1)$ symmetry, autocorrelation functions are expected to show hydrodynamic behaviour at late times, see e.g. \cite{dupont2020universal}. How does the latter combine with the effects of the SZMs on edge autocorrelators?
\item{} Weakly breaking integrability in models with SZMs gives rise to ``almost strong" zero modes \cite{kemp2017long,else2017prethermal}, which lead to edge coherence times that remain exponentially long in the perturbing couplings. How does the presence of a $U(1)$ symmetry and the associated hydrodynamic modes affect this physics?
\end{itemize}

We show how the behaviour in $\mathds{Z}_2$-broken cases is distinct.
To see how a normalizable ESZM operator $\Psi$ with support around the left boundary affects $C^{\cal O}(t)$ from \eqref{autocorr}, we choose the operator ${\cal O}(0)$ to transform under symmetries in the same fashion as $\Psi$. A natural choice, the one we make below, is ${\cal O}(0) = S_1^z$.  We then can decompose
${\cal O}(0)$ as
\be
{\cal O}(0)=c_1\Psi+c_2{\cal O}'\ ,\qquad {\rm Tr}\big(\Psi^\dagger{\cal O}'\big)=0\ ,\qquad c_1\neq 0\ ,
\ee
where $c_1={\cal O}(L^0)$ for large system sizes and ${\cal O}'$ is by construction localized around the left boundary. We define the norm of the ESZM via
\begin{align}
|\!|\Psi|\!|^2=(2S+1)^{-L}\,{\rm Tr}\left[\Psi^\dagger\Psi\right]=1
\label{normdef}
\end{align}
Using this norm then yields
\be
C^{\cal O}_0(t)=|c_1|^2+|c_2|^2C^{{\cal O}'}_0(t)\ .
\ee
The autocorrelator $C^{{\cal O}'}(t)$ is expected to decay in time to a value that vanishes as $L\to\infty$, which in turn implies that $C^{\cal O}(t)$ decays to a finite value $|c_1|^2$ set by the overlap of ${\cal O}(0)$ with $\Psi$. The behaviour of autocorrelation functions in presence of SZMs is different: a new time scale $t^*(L)$ that grows with $L$ arises because SZMs only approximately commute with the Hamiltonian. In models with $\mathds{Z}_2$ symmetry \cite{kemp2017long} this results in $C^{\cal O}(t)$ displaying oscillatory behaviour or slow decay for $t>t^*(L)$. 

\subsection{Edge coherence in the spin-\sfix{$\tfrac12$} XXZ chain}
We start with the spin-$\tfrac12$ XXZ chain, with Hamiltonian given by \eqref{Hopenper},\eqref{hhalf} with 
\begin{align}
b^{(\tfrac12)}_1=\,0\ ,\qquad b^{(\tfrac12)}_L=h_L\sigma^z_L.
\end{align}
It has long been known that including this (or any) boundary field does not destroy the integrability here; for an elementary proof see \cite{FGVV}. When $h_L=0$, the $U(1)\rtimes \mathbb{Z}_2$ symmetry requires that all energy levels with non-zero magnetization automatically form pairs with eigenvalues $\pm s_z$. Thus to probe any non-obvious consequences of any SZM one needs only consider the contribution of the $S^z=0$ sector to the autocorrelation function, namely
\be
f_0(t)=\frac{1}{C^{L}_{L/2}}\sum_n\big\langle E_n,s^z=0|\sigma^z_1(t)\sigma^z_1(0)|E_n,s^z=0\big\rangle\ .
\label{autocorrsz=0}
\ee
Here $C^n_m$ denotes a binomial coefficient. Even when breaking this pairing by taking $h_L\ne 0$, the projected autocorrelator $f_0(t)$ dominates the infinite-temperature autocorrelator.

\begin{figure}[ht]
\centering
\includegraphics[scale=0.35]{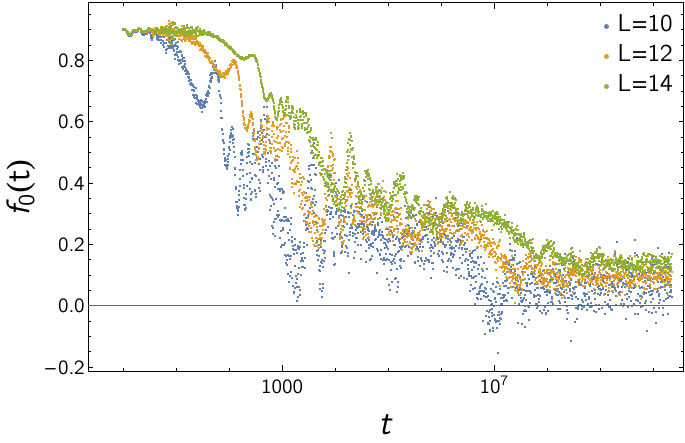}
\caption{\small Projected infinite-temperature autocorrelation function $f_0(t)$ for
  $S=\tfrac12$, $\Delta=4.5$, $h_L=0$ and system sizes $L=10,12,14$. }
\label{fig:autoxxz_hl0}
\end{figure}

We start with free boundary case $h_L=0$, which possesses a SZM \cite{fendley2016strong}. Results for the projected autocorrelation function $f_0(t)$ for three different system sizes are shown in Fig.~\ref{fig:autoxxz_hl0}.
We observe that at early times $f_0(t)$ settles to a plateau up to a time $t^*(L)$ that grows with system size, and then exhibits a very slow, possibly logarithmic, decay in time. This is markedly different from the effects due to SZMs in systems with discrete symmetries \cite{kemp2017long}, where the edge autocorrelator first settles to a plateau up a time $t^*(L)$ and then either displays oscillatory behaviour or exponential decay. A possible scenario for explaining this difference in the regime $t>t^*(L)$ is hydrodynamic behaviour arising from the conserved $U(1)$ charge in the spin-$\tfrac12$ XXZ chain, see e.g.\cite{dupont2020universal}. 

\begin{figure}[ht]
\centering
\includegraphics[scale=0.3]{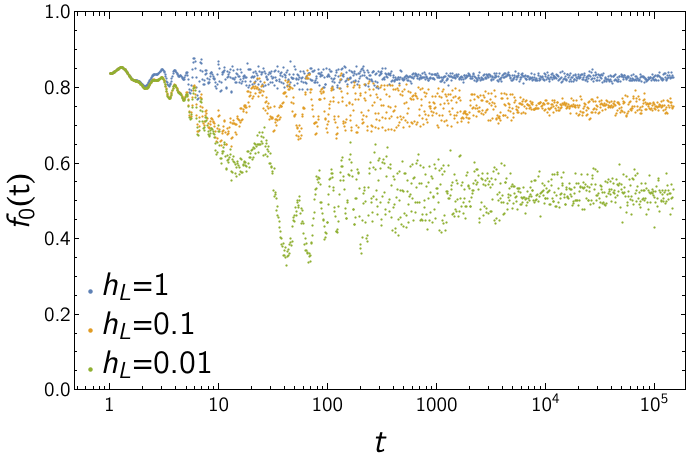}
\includegraphics[scale=0.27]{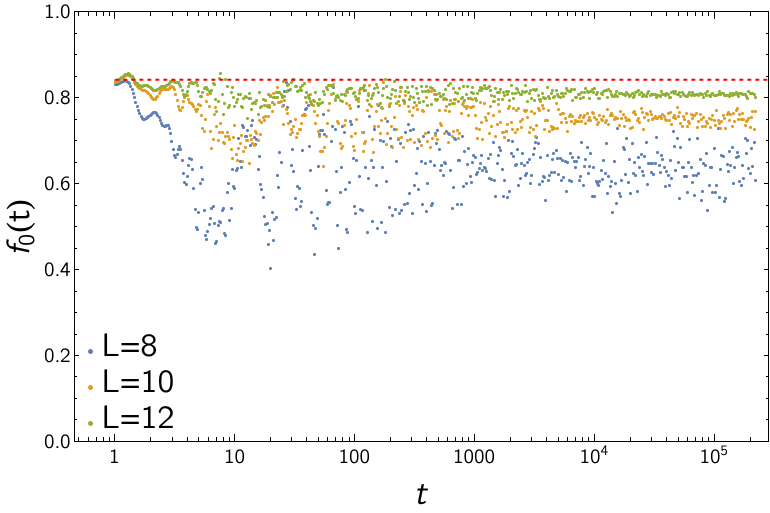}
\caption{\small Left panel: Projected infinite-temperature autocorrelation function $f_0(t)$ for
  $S=\tfrac12$, $\Delta=3.5$, system size $L=10$ and several values of $h_L$. At late times $f_0(t)$ relaxes to a finite constant due to the presence of an ESZM. Right panel: same for $h_L=0.1$ and three different system sizes. The dashed red line is the prediction from the ESZM projection \fr{inequality} for $L\gg 1$.}
\label{fig:autoxxz_hl}
\end{figure}

For $h_L\neq 0$ we have an ESZM, which leads to the behaviour shown in Fig.~\ref{fig:autoxxz_hl}. After some initial decay the autocorrelation function settles at a finite, non-zero value, which is determined by the overlap of $\sigma^z_1$ with the ESZM operator, as previously discussed.
The left panel shows that as $h_L$ becomes small and we approach the case in which we have a SZM rather than a conserved charge, an oscillatory crossover behaviour in $f_0(t)$ occurs at intermediate times. At very late times the amplitude of the oscillations is seen to decrease. In the right panel of Fig.~\ref{fig:autoxxz_hl} we show that for the small system sizes considered here the plateau reached at late times varies significantly with $L$, but eventually settles to a value that is independent of $h_L$ and compatible with 
the arguments based on the overlap of $\sigma^z_1$ with the ESZM operator presented in \emph{cf.} section \ref{ssec:locality}.
\subsection{Edge coherence in the perturbed spin-\sfix{$\tfrac12$} chain
}
\begin{figure}[ht!]
\centering
\includegraphics[scale=0.27]{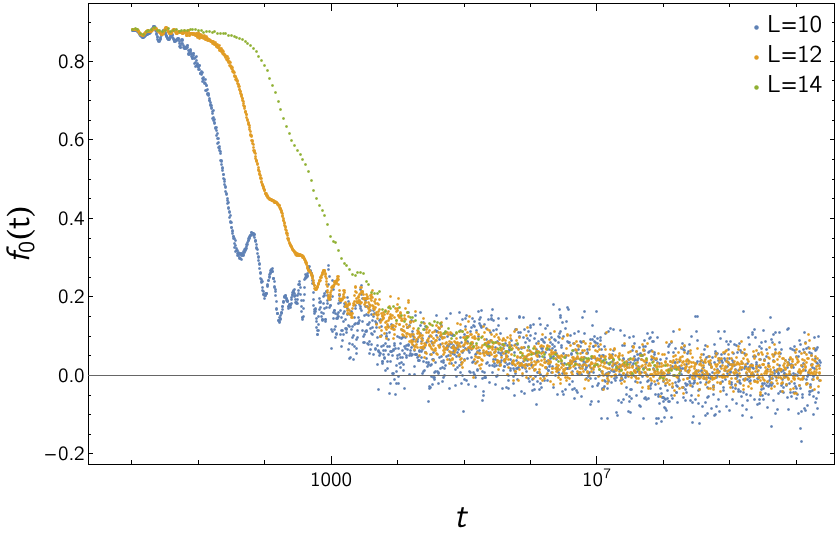}
\caption{\small Projected infinite-temperature autocorrelation function $f_0(t)$ for
  $S=\tfrac12$, $\Delta=4.5$, $h_L=0$, $J^x_2=0.25$ and system sizes $L=10,12,14$. We see the emergence of plateau at intermediate times, the duration of which grows with $L$.}
\label{fig:autoxxz_hl0_pert}
\end{figure}

We now consider the effects of adding an integrability breaking perturbation $\delta H$ to the spin-$\tfrac12$ chain that preserves the $U(1)$ symmetry, namely
\be
\delta H=J_2^x\sum_{j=1}^{L-2}\sigma^x_{j}\sigma^x_{j+2}+\sigma^y_{j}\sigma^y_{j+2}\ .
\label{deltaH2}
\ee
For the $\mathbb{Z}_2$-preserving case $h_L$\,=\,0, we show results for $f_0(t)$ in Fig.~\ref{fig:autoxxz_hl0_pert}. 
The most notable feature is the plateau at intermediate times, the duration of which grows with $L$. At very late times the projected autocorrelator decays to a very small value. Both features are in agreement with the expectation of an ``almost SZM" \cite{kemp2017long,else2017prethermal} for $h_L=0$, showing the $U(1)$ symmetry here does not affect the edge physics.

\begin{figure}[ht]
\centering
\includegraphics[scale=0.3]{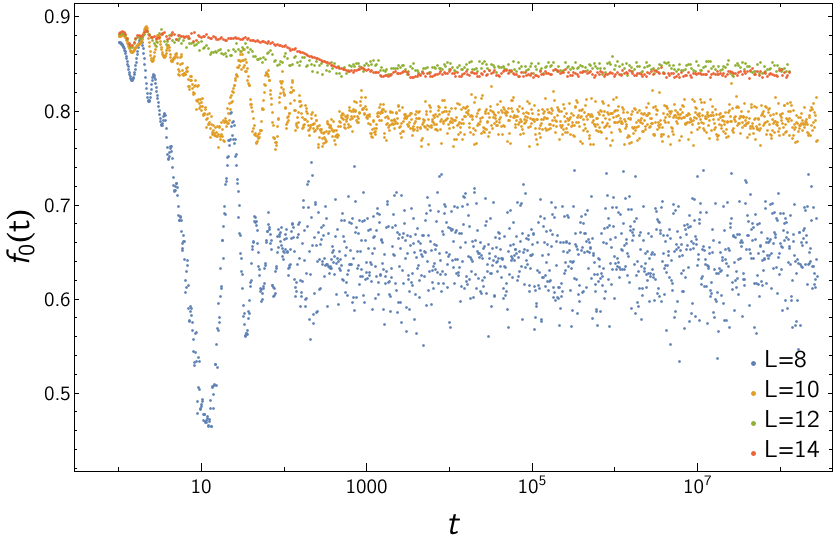}
\caption{\small Projected infinite-temperature autocorrelation function $f_0(t)$ for
  $S=\tfrac12$, $\Delta=4.5$, $h_L=0.1$, $J^x_2=0.25$ and system sizes $L=10,12,14$.}
\label{fig:autoxxz_hl01_pert}
\end{figure}

In Fig.~\ref{fig:autoxxz_hl01_pert} we show the corresponding results for $h_L=0.1$. In this case the perturbation \fr{deltaH2} breaks an ESZM.
We see that there again is a plateau, the duration of which grows with $L$. Moreover, at large times, the projected autocorrelator now does not decay to zero but a finite value, at least at accessible system sizes.  However, given that the value for $L$\,=14 is slightly lower than that for $L$\,=12, this second plateau could be  a small-size artefact. Nevertheless, these results shows a marked difference in the effects of an integrability-breaking perturbation on ESZMs as compared to SZMs. 

\subsection{Edge autocorrelations in spin-1 chains}
\begin{figure}[ht]
\centering
\includegraphics[scale=0.27]{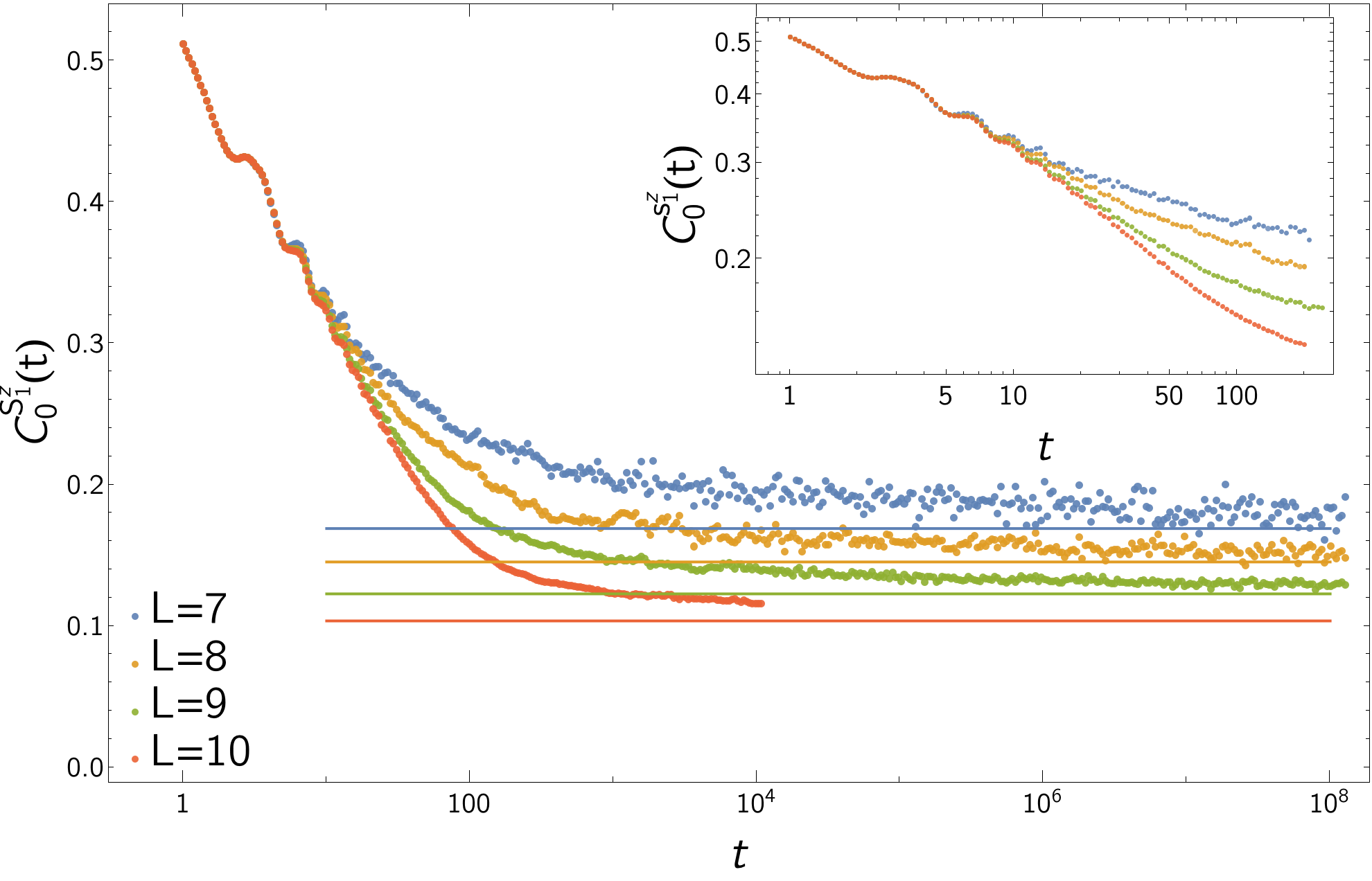}
\caption{\small Infinite-temperature edge autocorrelation function $C_0^{s^z_1}(t)$ for the
non-integrable, open $S=1$ XXZ chain \eqref{nonintXXZ} with $\Delta=3.7622$ and system sizes $L=6,8,10$. The solid lines are the infinite time averages \fr{timeaverage}. The inset shows early times on a double logarithmic scale.}
\label{fig:S1XXZ}
\end{figure}

We now turn to infinite-temperature autocorrelation functions in spin-1 chains. To provide some
helpful context, we start with the spin-1 XXZ chain with open boundary conditions
\be
H^{(S=1)}_{\rm XXZ}=\sum_{j=1}^{L-1}s^x_js^x_{j+1}+s^y_js^y_{j+1}+\Delta s^z_js^z_{j+1}\ .
\label{nonintXXZ}
\ee
This model is not integrable, but it is $U(1)$ symmetric and therefore will exhibit hydrodynamic tails in autocorrelation functions. In Fig.~\ref{fig:S1XXZ} we show the infinite-temperature edge autocorrelation function for system sizes $L=6,8,10$.
We observe a very slow (possibly power law) decay in time, followed by a plateau, the value of which depends on $L$. A useful indicator of the late-time behaviour at fixed $L$ are the infinite time averages
\begin{align}
\overline{C_0^{s^z_1}}(L)=\lim_{T\to\infty}\int_0^T dt\ C_0^{s^z_1}(t)
=\frac{1}{3^L}\sum_{s=-L}^{L}\sum_{n,m}|\langle
E_n,s|s^z_1|E_m,s\rangle|^2\delta_{E_n,E_m}\ ,
\label{timeaverage}
\end{align}
where $|E_n,s\rangle$ are energy eigenstates with $s^z$ eigenvalue $s$.
\subsection{Edge coherence in the integrable spin-1 chain}
\begin{figure}[ht]
\centering
\includegraphics[scale=0.33]{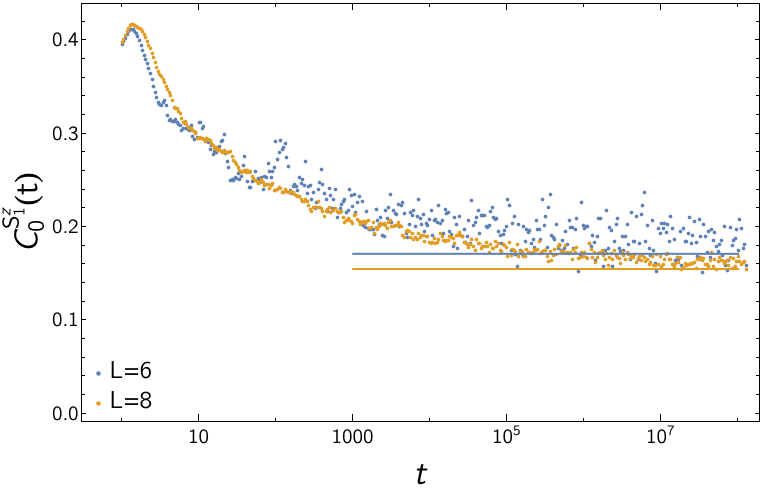}
\caption{\small Infinite-temperature edge autocorrelation function $C^{s^z_1}(t)$ for the $S=1$ chain with Hamiltonian density \fr{hspin1} with $\eta=1$, vanishing boundary fields $b^{(1)}_1=0=b^{(1)}_L$ and system sizes $L=6,8$. The solid lines are the infinite-time averages \fr{timeaverage}. }
\label{fig:S1nonintBCs}
\end{figure}

We now turn to the integrable spin-1 chain \fr{Hopenper}, \fr{hspin1}. We first consider open boundary conditions without any boundary fields:  $b^{(1)}_1=0=b^{(1)}_L$ .
As opposed to the spin-$\tfrac12$ case, these boundary conditions are {\em not} compatible with integrability \cite{mezincescu1990bethe}. In Fig.~\ref{fig:S1nonintBCs} we show the infinite-temperature edge autocorrelation function for system sizes $L=6,8$. 
We see that the edge autocorrelator behaves in a qualitatively similar way to the spin-1 XXZ chain. In particular there is no evidence suggesting the presence of a SZM for the non-integrable boundary conditions considered.
\subsubsection{Integrable, spin-flip invariant boundary conditions I}
\begin{figure}[ht]
\centering
\includegraphics[scale=0.32]{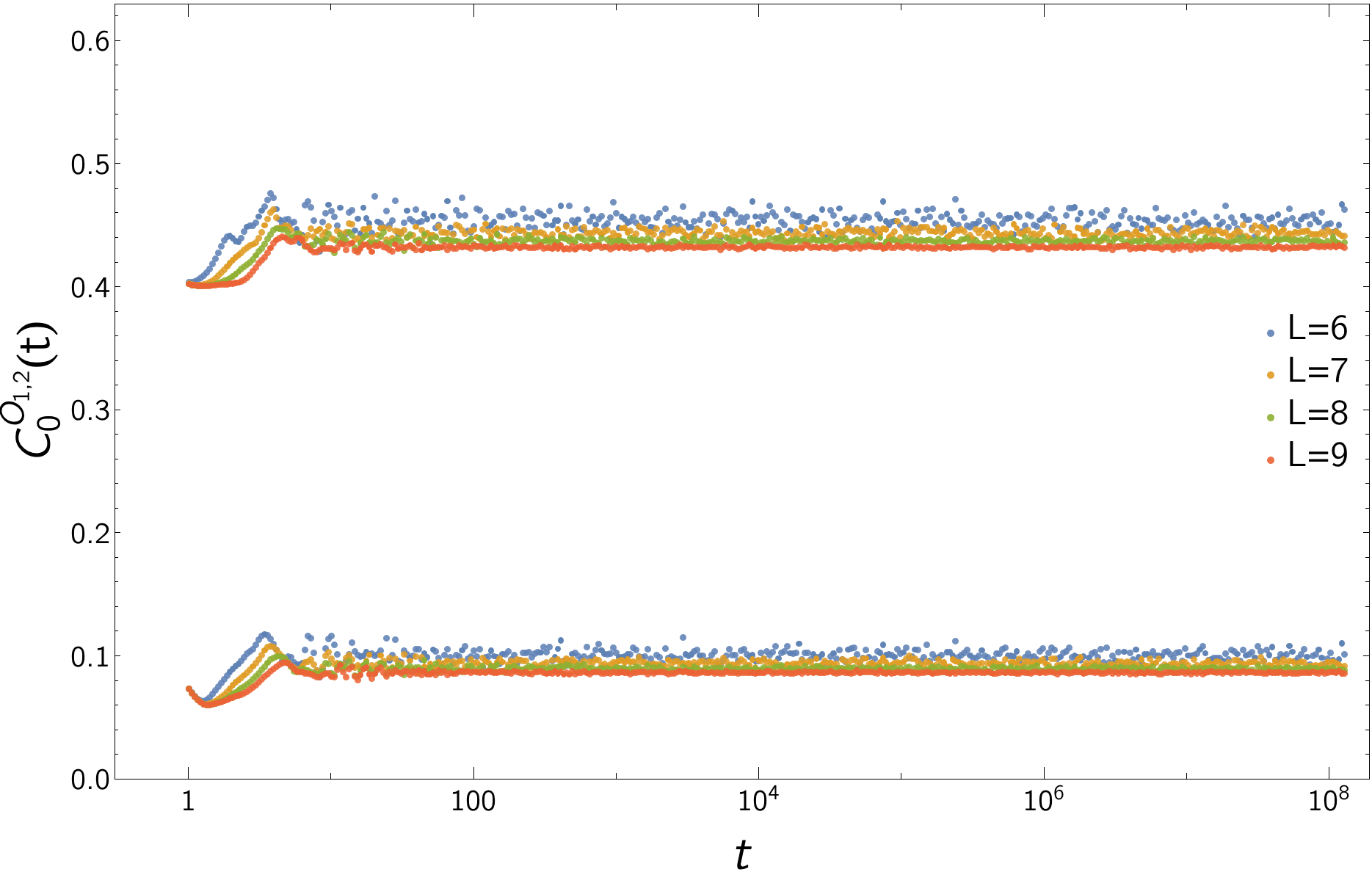}
\vspace{-.15in}
\caption{\small  Infinite-temperature autocorrelation functions \fr{autocorr} for ${\cal O}_1=s^z_1$ (upper two curves) and ${\cal O}_2=(1-(s^z_1)^{2})s^z_2$ (lower two curves) in the integrable open spin-1 chain with $\eta=1$, boundary potentials \fr{bpot}, and system sizes $6\leq L\leq9$. These results suggest the presence of an ESZM operator localized around the left boundary.}
\label{fig:autocorr}
 \end{figure}

We now fix the boundary terms in the integrable spin-1 Hamiltonian to be
\be
b^{(1)}_1=\big(1+J_z-\sqrt{2(1+J_z)}\big)(S^z_1)^2\ ,\qquad
b^{(1)}_L=\big(1+J_z+\sqrt{2(1+J_z)}\big)(S^z_L)^2.
\label{bpot}
\ee
These boundary conditions are integrable \cite{mezincescu1990bethe} (in the paramerization used in section \ref{sec:ESZM} these correspond to $\xi^+=i\pi/2$, $\xi^-=0$) and spin-flip invariant. As shown below, they lead to the presence of an ESZM and an exact two-fold degeneracy of energy eigenstates in the $s^z=0$ sector. 
The behaviour of infinite-temperature autocorrelation functions $C_0^{s^z_1}(t)$ and ${\cal O}_2=(1-(s^z_1)^{2})s^z_2$ for these boundary conditions is shown for $\eta=1$ in Fig.~\ref{fig:autocorr}.
Both autocorrelation functions in Fig.~\ref{fig:autocorr} approach non-zero stationary values at late times. 
The comparison between different lattice lengths shows that finite-size effects are already small, and the observed plateau values are close to the infinite-time averages. Plotting the plateau values for the different finite sizes against $1/L$ shows a linear dependence, and extrapolating to $L\rightarrow\infty$ gives finite values for the late-time behaviour of both autocorrelation functions shown in Fig.~\ref{fig:autocorr}. Changing the integrable boundary conditions at site $L$ leads to the same kind of behaviour. These observations are clear fingerprints of the presence of an ESZM localized around the left boundary. Interestingly, in contrast to the $S=\tfrac12$ case, this requires the presence of boundary potentials at both boundaries. 

Taking the boundary fields to be identical and of the form
\be
b^{(1)}_1=\big(1+J_z-\sqrt{2(1+J_z)}\big)(S^z_1)^2\  ,\qquad
b^{(1)}_L=\big(1+J_z-\sqrt{2(1+J_z)}\big)(S^z_L)^2\ .
\label{bpot2}
\ee
also preserves the integrability \cite{mezincescu1990bethe} (in the parametrization used in section \ref{sec:ESZM} these correspond to $\xi^+=\xi^-=i\pi/2$). However, changing the field at the far edge changes the ESZM into an SZM as is shown below. In Fig.~\ref{fig:autocorrS1_SZM} 
we show infinite-temperature autocorrelation function of $s^z_1$ 
with $\eta=1$, boundary potentials \fr{bpot2}, and system sizes $6\leq L\leq 9$.
We observe that the late-time behaviour is similar to the one for non-integrable boundary conditions shown in Fig.~\ref{fig:S1nonintBCs}. However, at times $1\leq t\leq 10$ (for the lattice lengths considered here) we see that $C_0^{S^z_1}(t)$ relaxes to a plateau, the width of which grows with system size. This early-time behaviour is similar to that of the spin-$\tfrac12$ case in presence of a SZM,
\emph{cf.} Fig.~\ref{fig:autoxxz_hl0}.

\begin{figure}[!ht]
\centering
\includegraphics[scale=0.32]{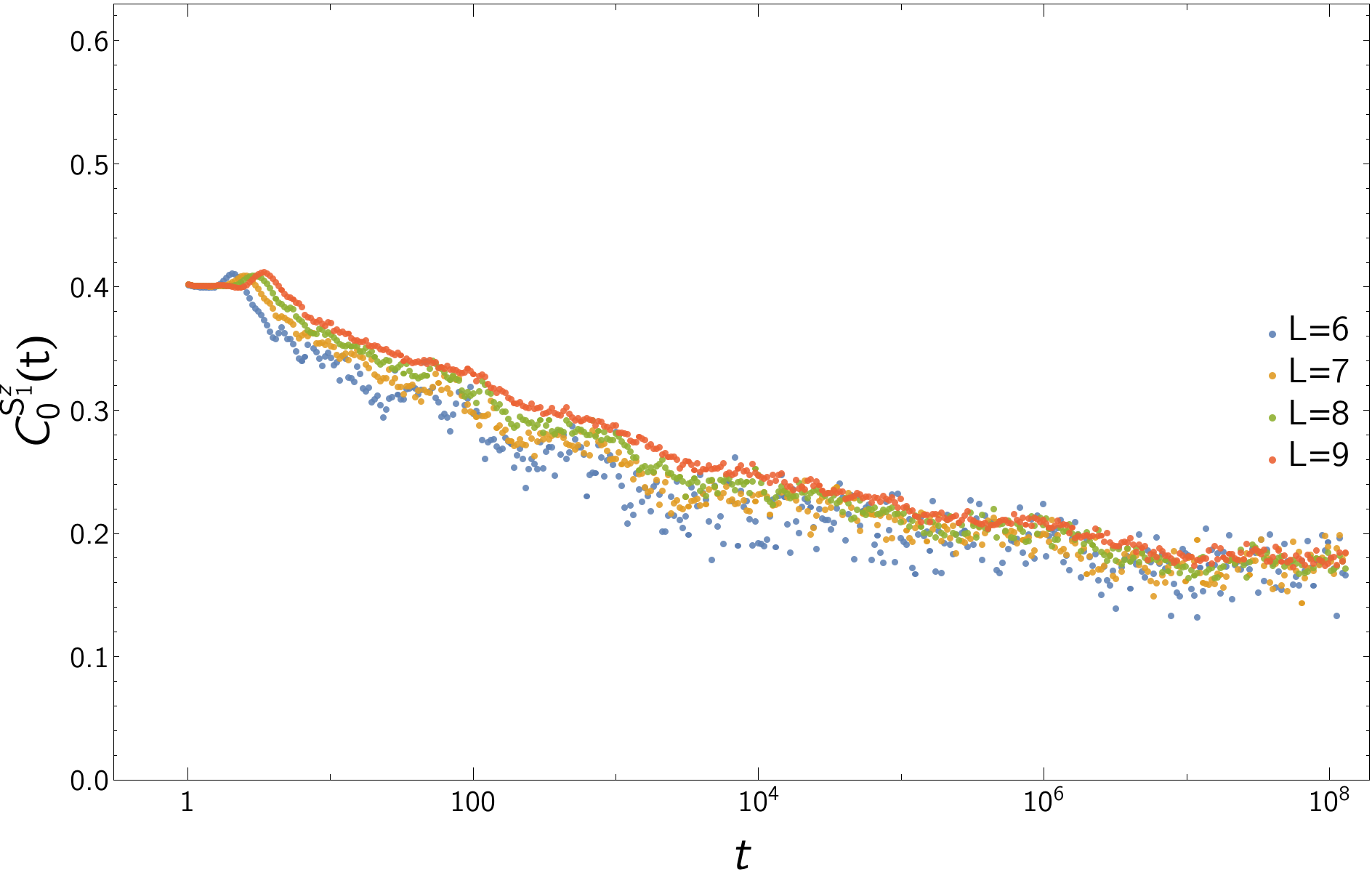}
\vspace{-.15in}
\caption{\small Infinite-temperature autocorrelation functions \fr{autocorr} for ${\cal O}_1=s^z_1$ in the integrable open spin-1 chain with $\eta=1$, boundary potentials \fr{bpot2}, and system sizes $6\leq L\leq 9$. These results suggest the presence of a SZM operator localized around the left boundary. }
\label{fig:autocorrS1_SZM}
 \end{figure}

\subsubsection{Integrable, spin-flip invariant boundary conditions II}
\begin{figure}[ht]
\centering
\includegraphics[scale=0.32]{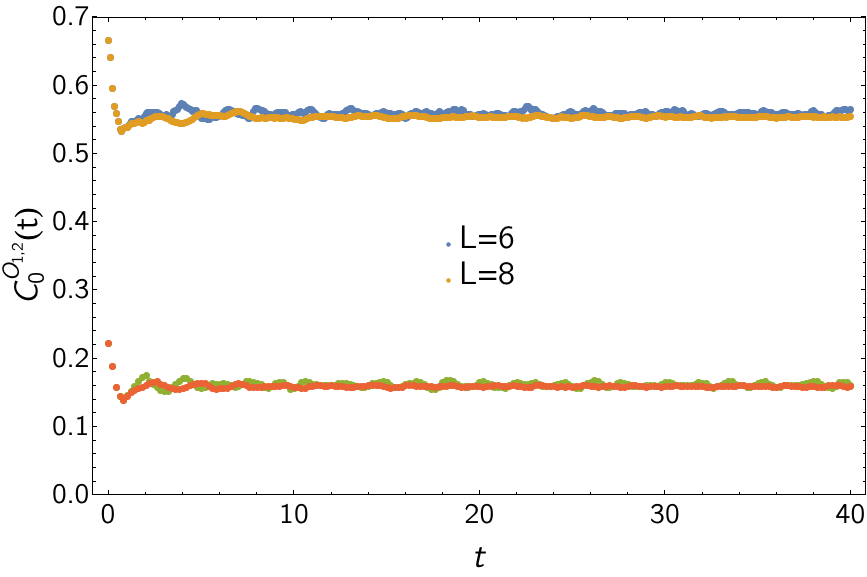}
\vspace{-.15in}
\caption{\small  Infinite-temperature autocorrelation functions \fr{autocorr} for ${\cal O}_1=s^z_1$ (upper two curves) and ${\cal O}_2=(1-(s^z_1)^{2})s^z_2$ (lower two curves) in the integrable open spin-1 chain with $\eta=1$, boundary potentials \fr{bpot3}, and system sizes $L=6$ and $L=8$. }
\label{fig:autocorr_x}
 \end{figure}

\begin{figure}[!ht]
\centering
\includegraphics[scale=0.32]{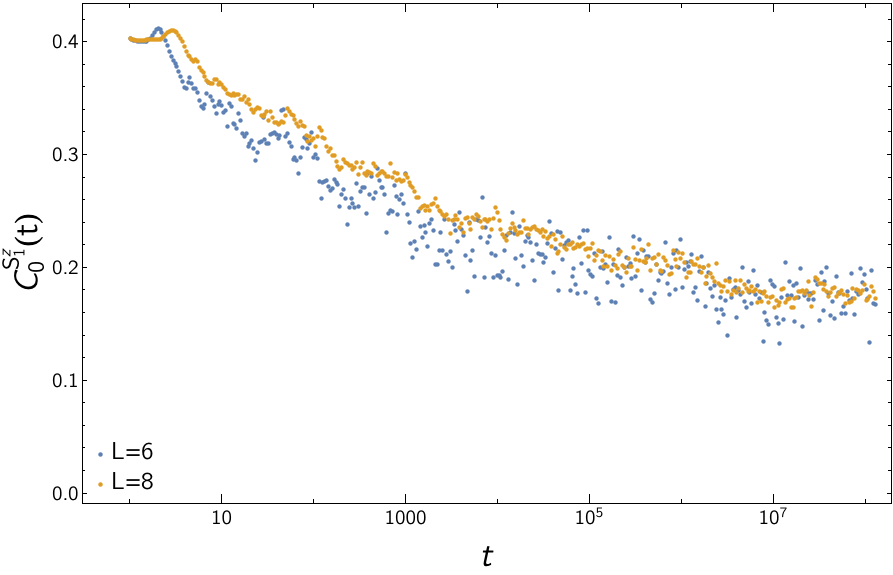}
\vspace{-.15in}
\caption{\small Infinite-temperature autocorrelation functions \fr{autocorr} for ${\cal O}_1=s^z_1$ in the integrable open spin-1 chain with $\eta=1$, boundary potentials \fr{bpot4}, and system sizes $L=6$ and $L=8$. }
\label{fig:autocorrS1_SZM_x}
 \end{figure}

Spin-flip invariant boundary conditions are also obtained by flipping the two boundaries
(in the parametrization used in section \ref{sec:ESZM} these correspond to $\xi^+=0$, $\xi^-=i\pi/2$)
\be
b^{(1)}_1=\big(1+J_z+\sqrt{2(1+J_z)}\big)(S^z_1)^2\  ,\qquad
b^{(1)}_L=\big(1+J_z-\sqrt{2(1+J_z)}\big)(S^z_L)^2.
\label{bpot3}
\ee
The infinite-temperature autocorrelation functions $C_0^{s^z_1}(t)$ and ${\cal O}_2=(1-(s^z_1)^{2})s^z_2$ for these boundary conditions are shown for $\eta=1$ in Fig.~\ref{fig:autocorr_x}.
 Both autocorrelation functions approach non-zero stationary values at late times, and finite-size effects are seen to be very small. This suggests the presence of an ESZM localized around the left boundary.
 Finally we consider identical boundary potentials of the form 
\be
b^{(1)}_1=b^{(1)}_L=\big(1+J_z+\sqrt{2(1+J_z)}\big)(S^z_1)^2\ .
\label{bpot4}
\ee
In the parametrization used in section \ref{sec:ESZM} these correspond to $\xi^+=0=\xi^-$.
The infinite-temperature autocorrelation function $C^{s^z_1}_0(t)$ for these  boundary conditions is shown in Fig.~\ref{fig:autocorrS1_SZM_x}. 
The behaviour is quite similar to the one for boundary potentials \fr{bpot2} shown in Fig.~\ref{fig:autocorrS1_SZM}, and suggests the presence of a SZM operator localized around the left boundary.
\subsection{Edge coherence in the perturbed integrable spin-1 chain}
Finally we examine the behaviour of edge autocorrelators after  perturbing away from this integrable spin-1 chain with an ESZM. To that end we add an integrability-breaking single-ion anisotropy term to the model $D$ as in \eqref{HD}. 
\begin{figure}[ht]
\centering
\includegraphics[scale=0.35]{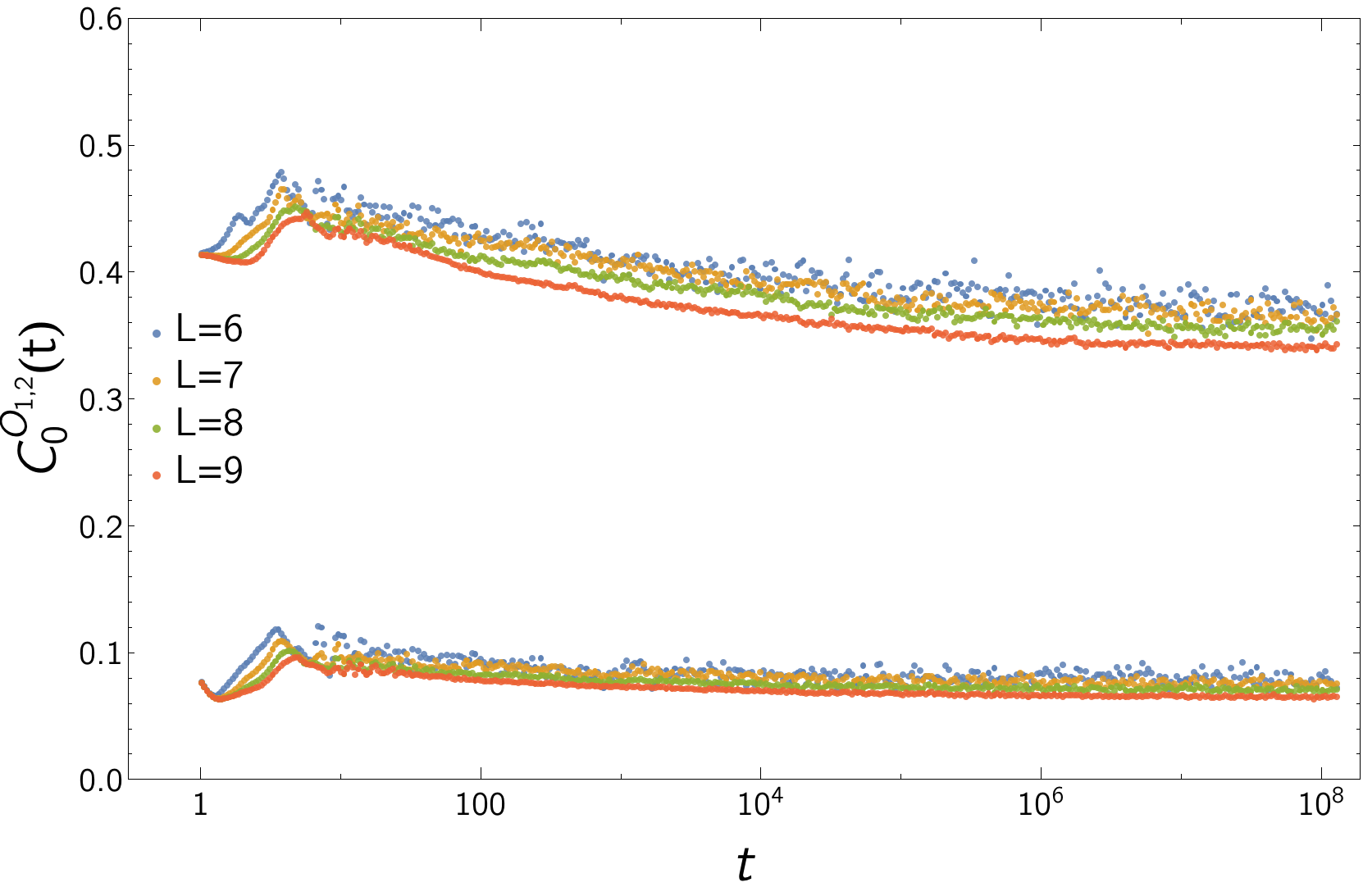}
\caption{\small Infinite-temperature edge autocorrelation function $C^{{\cal O}_{1,2}}(t)$ for 
${\cal O}_1=s^z_1$ (upper two curves) and ${\cal O}_2=(1-(s^z_1)^{2})s^z_2$ in the integrable spin-1 chain with $\eta=1$, boundary potentials \fr{bpot} perturbed by \fr{HD} with $D=0.25$ and system sizes $6\leq L\leq 9$.}
\label{fig:S1D025}
\end{figure}
Results for the autocorrelation function of 
${\cal O}_1=s^z_1$ and ${\cal O}_2=(1-(s^z_1)^{2})s^z_2$ are shown in Fig.~\ref{fig:S1D025} for system sizes $6\leq L\leq 9$ and values $D=0.25$.
We see that as expected the autocorrelators decay more quickly, but they
still appear to approach constants at late times. How these plateau values evolve with larger system sizes $L$ at fixed $D$ is an interesting question, but one not possible to address using our numerical approach.

\section{Exact strong zero modes from transfer matrices}
\label{sec:ESZM}

We explained in section \eqref{sec:spin1intro} that the structure of SZMs in the integrable spin-1 chain cannot be as elegant as for the spin-$\tfrac12$ case. Since for open boundary conditions and a $\mathbb{Z}_2$ symmetry, it has three ground states exponentially close in energy, making impossible a perfect pairing of states. Nonetheless, the numerical analysis of the edge autocorrelation function in the preceding section strongly suggests that the spin-1 chain still exhibits very special behaviour near the edge, much of it reminiscent of what happens at spin $\tfrac12$. 

To understand this behaviour, in this section we construct operators for spin-$S$ integrable XXZ Hamiltonians that strongly resemble exact strong zero modes. They commute exactly with the Hamiltonian, and are built in a fashion naturally generalising that for the spin-$\tfrac12$ chain. Nonetheless, they do {\em not} satisfy the full properties typically demanded of an SZM or ESZM. They do not square to the identity, and they are not perfectly localised near the edge. However, we will show both of these properties do hold true in certain operator norms. Thus they greatly resemble their spin-$\tfrac12$ analogs, and so will still call them ESZMs. 

The key to our construction comes from several observations of Ref.~\cite{FGVV} regarding the strong zero mode for the spin-$\tfrac12$ chain. There it was shown that the corresponding SZM can be written as a matrix-product operator, and extended further to give a generating function of conserved charges.  Crucially, this generating function of written as a product of transfer matrices of an associated classical problem, in that case the eight-vertex model. Using the standard techniques of integrability, one finds that such transfer matrices generate a family of commuting quantities by construction. Moreover, by a procedure called fusion, one can construct commuting transfer matrices for integrable higher-spin models. 

Such transfer matrices automatically commute with the Hamiltonian of higher-spin chains. One then can define ESZMs for these higher-spin chains by taking the analogous limit to the spin-$\tfrac12$ case. However, while the formalities go through nicely, some of the physical properties of the general ESZMs are not as simple as those for spin $\tfrac12$. For example, we explain how at least for spin 1, the ESZM is localized at the edge only in the sense of an operator norm.

\subsection{Commuting family of transfer matrices for spin \sfix{$S$}}
We proceed by introducing a family of two-dimensional classical integrable vertex models. For a given physical spin $S$ a well-understood construction \cite{ZHOU1995619,Mezincescu_1992,ZHOU1996504} yields a family of transfer matrices $T^{(S,S')}(u)$ labeled by a spectral parameter $u\in\mathds{C}$ and an auxiliary spin $S'=0,\tfrac12,1, \ldots$. These transfer matrices have the essential property that they commute for any two pairs of auxiliary spins and spectral parameters:
\be
[T^{(S,S_1)}(u),T^{(S,S_2)}(v)]=0\ .
\label{eq:commutationtransfermatrices}
\ee
The Hamiltonian $H^{(S)}$ of the integrable spin-$S$ chain is obtained by taking derivatives of $T^{(S,S)}(u)$ at a special value of $u$ and therefore automatically commutes with the entire family of transfer matrices.
Much of the following discussion will be concerned with the simplest case of auxiliary spin $S'=\tfrac12$, while the case of higher auxiliary spin will be briefly discussed in section \ref{sec:higherauxspin}. 

The basic building blocks of the transfer matrices are the L-operators ${\cal L}^{(S,S')}_j(u)$ that act on the tensor product $\mathds{C}^{2S'+1}\otimes\mathds{C}^{2S+1}$ of auxiliary and ``physical" spaces. These operators encode the local interactions of the model, and can be treated as $(2S'+1)\times(2S'+1)$ matrices with each entry acting non-trivially on the $j$th spin.  They can be pictured as
\begin{center}
 \begin{tikzpicture}
\draw[blue, line width=1] (0.5,0) -- (1.5,0);
\draw[line width=1,>-,>=latex] (0.5,0) -- (0.6,0);
\draw[red,line width=1,>-,>=latex] (1,-0.5)  -- (1,0.5);
\draw[line width=1, rounded corners=2pt] (1-0.2,0)--  (1-0.15,-0.15)-- (1,-0.2) ;
\node at (1,-0.75) {$j$};\node at (0.75,-0.25) {\small $u$}; 
\node at (-1,0) {${\cal L}_j^{(S,S')}(u)=$};
\end{tikzpicture}
\end{center}
The single-row transfer matrices constructed from the L-operators are
\begin{align}
 T_a(u) &= \mathcal{L}_{1}^{(S,S')}(u) \ldots \mathcal{L}^{(S,S')}_{L}(u) \ ,\qquad
\hat T_a(u) = \mathcal{L}^{(S,S')}_{L}(u) \ldots \mathcal{L}^{(S,S')}_{1}(u)\ ,
\end{align}
where $a$ labels the auxiliary space.

Following \cite{FGVV}, the transfer matrices $T^{(S,S')}(u)$ we utilise are double-layered, and of the form studied in \cite{sklyanin1988boundary}. They are conveniently pictured as in Fig.~\ref{fig:TMSSprime}.
\begin{figure}[ht]
\centering
 \begin{tikzpicture}
\draw[blue, line width=1, rounded corners=10pt] (0.5,0) -- (8,0) -- (8.5,0.5) -- (8,1) -- (0,1) -- (-0.5,0.5) -- (0,0) -- (0.5,0);
\draw[line width=1,>-,>=latex] (0.2,0) -- (0.3,0);
\node at (-1,0.5) {$K^+_{S'}(u)$};
\node at (9,0.5) {$K^-_{S'}(u)$};
\foreach \x in {1,3,...,7} { 
\draw[red,line width=1,>-,>=latex] (\x,-0.5)  -- (\x,1.25);
\draw[line width=1, rounded corners=2pt] (\x-0.2,0)--  (\x-0.15,-0.15)-- (\x,-0.2) ;
\draw[line width=1, rounded corners=2pt] (\x+0.2,1)--  (\x+0.15,0.85)-- (\x,0.8) ;
}
\node at (1,-0.75) {$1$}; \node at (0.75,-0.25) {\small $u$}; \node at (1.25,0.75) {\small $u$};
\node at (3,-0.75) {$2$};\node at (2.75,-0.25) {\small $u$};\node at (3.25,0.75) {\small $u$};
\node at (5,-0.75) {$\ldots$};
\node at (7,-0.75) {$L$};\node at (6.75,-0.25) {\small $u$};\node at (7.25,0.75) {\small $u$};
\end{tikzpicture}
  \caption{\small The transfer matrix $T^{(S,S')}(u)$ acting on $L$ sites. The physical spin $S$ are represented in red, and the auxiliary spin $S'$ in blue.}
 \label{fig:TMSSprime}
 \end{figure} 
The vertices are the L-operators, and the $K_{S'}^\pm(u)$ are \emph{reflection matrices} \cite{cherednik1984factorizing,de1994boundary} that act on the auxiliary space and encode the boundary conditions. The transfer matrices thus can be written as the trace over the $2S'+1$-dimensional auxiliary space, namely
 \begin{align} 
T^{(S,S')}(u) &= \mathrm{tr}_a\left( K_a^+(u) T_a(u) K_a^-(u) \hat{T}_a(u)  \right)\,.
\label{eq:TM}
\end{align} 
Integrability allows one to find L-operators and reflection matrices that yield transfer matrices satisfying the commutation relation \eqref{eq:commutationtransfermatrices} \cite{sklyanin1988boundary}. We give explicit examples below.


\subsubsection{Two-dimensional auxiliary space \sfix{$S'=\tfrac12$}}

Given a fixed value of the physical spin $S$, we first consider the simplest family of transfer matrices $T^{(S,\frac12)}$, where the auxiliary space is two-dimensional. 
These have the advantage that they can be expressed in a relatively simple form in terms of local operators $S_j^{z,+,-}$ 
Furthermore, they can be used to construct the higher transfer matrices $T^{(S,S')}$ through a procedure called ``fusion" \cite{ZHOU1995619,Mezincescu_1992,ZHOU1996504,Piroli_2017} (see Appendix \ref{app:Tsystem}). 
Although the full set of conserved charges for the spin-$S$ Hamiltonian requires utilising all $S'$ (as is well known for ``bulk'' local or quasilocal charges \cite{ilievski2015complete, ilievski2016quasilocal}), we will see that $T^{(S,\frac12)}$ already generates interesting families of conserved operators.

For auxiliary spin $S'=\tfrac{1}{2}$, the L-operator on site $j$ can be compactly written in terms of a triplet of operators $S_j^{z,+,-}$ acting non-trivially on the jth spin, taking the form 
\begin{align} 
\mathcal{L}_j(u) &=
\frac{1}{\sinh(u+\tfrac{\eta}{2} + S\eta)}
\left(
\begin{array}{cc}
\sinh(u+\tfrac{\eta}{2} + \eta S_j^z) & \sinh\eta S_j^-
\\
\sinh\eta S_j^+  & \sinh(u+\tfrac{\eta}{2} - \eta S_j^z) 
\end{array}
 \right) \,.
 \label{eq:Laxop}
\end{align}
For transfer matrices to commute at different spectral parameters as in \eqref{eq:commutationtransfermatrices}, the operators must obey the defining relations of the quantum group $U_q(sl_2)$\cite{saleur1990integrable}:
\be 
[S^z, S^\pm] = \pm S^\pm \,, \qquad [S^+,S^-] = \frac{q^{2S^z}-q^{-2S^z}}{q-q^{-1}} = \frac{\sinh(2\eta S^z)}{\sinh \eta} \ .
\ee 
Explicit representations for these operators are given in Appendix \ref{app:suq2spins}. For $S$\,=\,$\tfrac12$, they reduce to Pauli matrices via $S_j^z = \sigma_j^z/2$, $S_j^\pm = (\sigma_j^x \pm i \sigma_j^y)/{2}$, while for $S$\,=\,1, they are proportional to the generators \eqref{spin_op}. 
In this paper we consider only diagonal reflection matrices. For $S'=\tfrac12$ they are  \cite{de1994boundary}
\be
K^{\pm}(u) = K(u+ \tfrac{\eta}{2} \pm \tfrac{\eta}{2} ,\xi_\pm)
\,, \qquad 
K(u,\xi) = \left( \begin{array}{cc}  \sinh(\xi + u ) & 0 \\ 0 & \sinh(\xi - u) \end{array}  \right) \,. 
\ee 
Much of the following discussion will be concerned with the simplest case of auxiliary spin $S'=\frac12$, but the case of higher auxiliary spin will be briefly discussed in section \ref{sec:higherauxspin}. 

\subsubsection{Integrable \sfix{spin-$S$} Hamiltonians}
The Hamiltonian $H^{(S)}$ of the open spin-$S$ chain is obtained by taking derivatives in $u$ of the transfer matrix $T^{(S,S)}(u)$. It therefore automatically commutes with any transfer matrix $T^{(S,S')}$.  Expressions for general $S$ are given in Appendix~\ref{app:suq2spins}. In the simplest case  $S=\tfrac12$ we recover the open chain from \eqref{Hopenper} with $\Delta = \cosh\eta$, where the boundary fields are parametrized in terms of $\xi^\pm$ as
\be 
\begin{split}
b^{(\frac12)}_1=  \sinh \eta \coth\xi^+ \sigma^z_1
\,, \qquad b^{(\frac12)}_L = \sinh \eta \coth \xi^-\sigma^z_L \,.
\label{boundaryfieldshalf}
\end{split}
\ee 
We note that for $\xi_\pm = i \frac{\pi}{2}$, the boundary fields vanish and the Hamiltonian is invariant under the $\mathbb{Z}_2$ global spin flip along the $z$ axis.

For $S$\,=\,1, eqn \fr{H_S} is equivalent to the Hamiltonian limit of the Zamolodchikov and Fateev model \cite{zamolodchikov1980model}  with open boundaries \cite{mezincescu1990bethe}. The bulk Hamiltonian is built from \eqref{hspin1} and \eqref{coefficients}, while the integrable diagonal boundary fields here are
\begin{align}
b^{(1)}_1&=\sinh(2\eta)\frac{- \sinh(2\xi^+)  s_1^z + \sinh\eta (s_1^z)^2}{\cosh\eta - \cosh(2\xi^+)}\ ,\nn
b^{(1)}_L&= \sinh(2\eta)\frac{- \sinh(2\xi^-)  s_L^z + \sinh\eta (s_L^z)^2}{\cosh\eta - \cosh(2\xi^-)}\ .
\label{bulkhamiltonianspin1}
\end{align}
For each boundary, there are now two choices of the parameters $\xi^\pm$ which result in a $\mathbb{Z}_2$-invariant Hamiltonian, namely $\xi^\pm=0,i\pi/2$. 

The observations made for $S=1/2$ and $S=1$ extend to higher spin, as can be checked from the explicit form of the boundary couplings given in Appendix~\ref{app:suq2spins} : for integer spin there are two possible $\mathbb{Z}_2$-invariant choices at each end, corresponding to $\xi^\pm=0,i\pi/2$, while for half-odd spin only the latter choice is allowed, as $\xi^\pm=0$ results in a diverging boundary coupling.

\subsection{ESZM \sfix{$\Psi$} constructed from transfer matrices \sfix{$T^{(S,\frac12)}$}}
\label{ssec:TShalf}

We now turn to the construction of an exact strong zero mode operator from the transfer matrix $T^{(S,{1}/{2})}(u)$.
The procedure closely follows the $S=\tfrac12$ case considered in \cite{FGVV,vernier2024strong}, but here is formulated abstractly in terms of the quantum-group generators $S^z,S^+,S^-$ in the physical space, and therefore holds for any representation of the physical spin. Similarly to $S=\tfrac12$, to get an ESZM we fix the left-boundary parameter to the $\mathbb{Z}_2$-invariant point $\xi^+=i\frac{\pi}{2}$, while the right-boundary parameter $\xi^-$ can be arbitrary. The corresponding boundary K-matrices are
\be 
K_{\rm a}^+(u) = i \cosh(\eta+u) \sigma_{\rm a}^0
\,, \qquad
K_{\rm a}^-(u) = \cosh(u)\sinh(\xi^-) \sigma_{\rm a}^0 
+
\sinh(u)\cosh(\xi^-) \sigma_{\rm a}^z \,,
\label{eq:Ku}
\ee 
where the subscript is a reminder that these matrices act on the auxiliary space. 
As pointed out above, for integer $S$ another $\mathbb{Z}_2$-invariant boundary condition can be obtained by taking $\xi^+=0$: we will briefly discuss the construction of ESZM for that case in section \ref{sec:EZSMxipluszero}. 

The cornerstone of the construction is the existence of a special value $u^*=i\tfrac{\pi}2$ of the spectral parameter where the very useful relation
\be 
\mathcal{L}_j\big(u^*\big)\, \sigma_{\rm a}^z\, \mathcal{L}_j\big(u^*\big) = \sigma_{\rm a}^z
\label{eq:LzL}
\ee 
holds, which can be depicted pictorially as
\begin{center}
 \begin{tikzpicture}
\draw[blue, line width=1, >-,>=latex,rounded corners=10pt] (6,0) -- (8,0) -- (8.5,0.5) -- (8,1) -- (6,1);
\node at (8.75,0.5) {$\sigma^z$};
\foreach \x in {7} { 
\draw[red, line width=1,>-,>=latex] (\x,-0.5)  -- (\x,1.25);
\draw[line width=1, rounded corners=2pt] (\x-0.2,0)--  (\x-0.15,-0.15)-- (\x,-0.2) ;
\draw[line width=1, rounded corners=2pt] (\x+0.2,1)--  (\x+0.15,0.85)-- (\x,0.8) ;
}
\node at (6.55,-0.25) {\small $\frac{i\pi}2$};\node at (7.45,0.75) {\small  $\frac{i\pi}2$};
\node at (10,0.5) {$=$};
 \begin{scope}[xshift=4cm]
\draw[blue, line width=1, >-,>=latex,rounded corners=10pt] (7,0) -- (8,0) -- (8.5,0.5) -- (8,1) -- (7,1);
\node at (8.75,0.5) {$\sigma^z$};
\foreach \x in {10} { 
\draw[red, line width=1,>-,>=latex] (\x,-0.5)  -- (\x,1.25);
}
\end{scope}
\end{tikzpicture}
 \end{center} 
One important consequence is that the transfer matrix $T^{(S,\frac12)}(u^*)$ vanishes. This follows by repeated application of \fr{eq:LzL} starting at site $L$ in Fig.~\ref{fig:TMSSprime} and moving leftwards, so that
\be
T^{(S,\frac12)}(i\pi/2)={\rm tr}_{\rm a}\Big[ K_{\rm a}^+\big(u^*\big)K_{\rm a}^-\big(u^*\big)\Big]\mathds{1}=0\ .
\label{TSvanish}
\ee


The proof of identity \eqref{eq:LzL} for arbitrary $S$ follows from explicit calculation starting with the L-operator defined in \eqref{eq:Laxop}. The off-diagonal entries of $\mathcal{L}_j(u^*)\, \sigma_{\rm a}^z\, \mathcal{L}_j(u^*)$ vanish, e.g.\ the 1,2 entry is proportional to 
\[S^- \cosh(\tfrac{\eta}{2}-\eta S^z) - \cosh(\tfrac{\eta}{2}+\eta S^z)S^-\ ,\]
which vanishes once one utilizes the explicit representation of quantum-generators given in \eqref{eq:Uqsl2}. The relation  \eqref{eq:LzL} then follows utilising the fact that the diagonal entries are proportional to the quadratic Casimir, as again can be verified using \eqref{eq:Uqsl2}.

%
%

One might expect the derivative ${T^{(S,\frac12)}}'(u^*)$ then to have nice properties. In the remainder of this section, we show that it is an ESZM in the sense described above. Namely, in addition to commuting with the Hamiltonian, it enjoys some nice locality properties. We make these properties precise in sections \eqref{sec:MPO} and ~\ref{ssec:locality} below. Here we give an explicit series expansion for this operator, showing how it does indeed naturally generalise the $S=\tfrac12$ case. 

The ESZM is thus defined to be proportional to
\begin{align}
{T^{(S,\frac12)}}'(u^*) 
=&  \mathrm{tr}_{\rm a} \left( {K_{\rm a}^+}'(u^*) T_{\rm a}(u^*) K_{\rm a}^-(u^*) \hat{T}_{\rm a}(u^*)  \right) 
+
 \mathrm{tr}_{\rm a} \left( K_{\rm a}^+(u^*) T_{\rm a}(u^*) {K_{\rm a}^-}'(u^*) \hat{T}_{\rm a}(u^*)  \right) \nonumber\\
&+ \frac{\mathrm{d}}{\mathrm{d}u}\left. \mathrm{tr}_{\rm a} \left( K_{\rm a}^+(u^*) T_{\rm a}(u) K_{\rm a}^-(u^*) \hat{T}_a(u)  \right) \right|_{u=u^*} 
\label{eq:expansionderivative}
\end{align} 
Since ${K^+}'(u^*)\propto K^+(u^*)$, 
the first term is proportional to $T^{(S,\frac12)}(u^*)$ and so vanishes. 
The last term can be written as a sum of terms where the derivative acts on an L-operator on site $j$, each of which can be simplified by repeatedly using
\eqref{eq:LzL} on sites $L$ to $j+1$, leaving in the identity operator on these sites. 
On site $j$ we use 
\begin{align}
\left. \frac{\mathrm{d}}{\mathrm{d}u}\big(\mathcal{L}_j(u) \sigma_{\rm a}^z \mathcal{L}_j(u) \big) \right|_{u=u^*}
&= 
\frac{\sinh\eta \cosh\eta}{\Delta_S^2}\left[ \sigma_{\rm a}^0  \Sigma^0_j 
+\sigma_{\rm a}^+ \Sigma^-_j + \sigma_{\rm a}^- \Sigma^+_j
+  \sigma_{\rm a}^z \Sigma^z_j   \right],
\end{align}
where we define $\Delta_S= \cosh(S\eta+\tfrac{\eta}{2})$  
and
\be 
\Sigma_j^\pm = -2 iS_j^\pm \frac{\sinh(\eta S_j^z \pm \tfrac{\eta}{2})}{\cosh\eta}  \,, \
 \Sigma_j^0 = \frac{\sinh(2\eta S_j^z)}{\sinh\eta} \,, \
\Sigma_j^z = 
 \frac{\cosh(2\eta S_j^z)}{\cosh\eta} - \frac{\sinh((2S+1)\eta)}{\sinh\eta\cosh\eta}.
\ee 
The matrices $\Sigma_j^\pm$ vanish for $S$\,=\,$\tfrac12$, greatly simplifying the expression of this ESZM.

To proceed, it is convenient to introduce a fictitious site $L+1$ and define
\be 
\Sigma_{L+1}^\pm = \Sigma_{L+1}^z = 0   \,, \qquad \Sigma_{L+1}^0 = \frac{\tanh\xi^- \Delta_S^2}{\cosh \eta \sinh\eta} \mathds{1}\ .
\label{SigmaLp1}
\ee
We then can combine the two remaining terms in \eqref{eq:expansionderivative} to get
\be
 {T^{(S,\frac12)}}'(u^*)
=  
\frac{\sinh^22\eta\cosh\xi^-  }{4i\Delta_S^2 \cosh\eta}\; 
\sum_{j=1}^{L+1} \sum_{\alpha\in\{0,+,-\}} \mathcal{T}_\alpha^{[\ldots j-1]} \Sigma_{j+1}^\alpha
\,,
\label{Treduced}
\ee
where we have introduced ``subsystem transfer matrices'' 
\begin{align} 
\mathcal{T}^{[\ldots j]}_\alpha &\equiv 
\mathrm{tr}_{a}(  \mathcal{L}_{1}(u^*) \ldots \mathcal{L}_{j}(u^*)   \sigma_{\rm a}^\alpha  \mathcal{L}_{j}(u^*) \ldots \mathcal{L}_{1}(u^*)  ) 
\label{eq:TMreduced}
\end{align}
which can conveniently pictured as 
\begin{center}
 \begin{tikzpicture}
\draw[blue,line width=1, rounded corners=10pt] (0.5,0) -- (6,0) -- (6.5,0.5) -- (6,1) -- (0,1) -- (-0.5,0.5) -- (0,0) -- (0.5,0);
\draw[line width=1,>-,>=latex] (0.2,0) -- (0.3,0);
\node at (6.7,0.5) {$\sigma^\alpha$};

\foreach \x in {1,3,5.5} { 
\draw[red,line width=1,>-,>=latex] (\x,-0.5)  -- (\x,1.25);
\draw[line width=1, rounded corners=2pt] (\x-0.2,0)--  (\x-0.15,-0.15)-- (\x,-0.2) ;
\draw[line width=1, rounded corners=2pt] (\x+0.2,1)--  (\x+0.15,0.85)-- (\x,0.8) ;
}
 \node at (0.5,-0.35) {\small $u^*$}; \node at (1.5,0.65) {\small $u^*$};
\node at (2.5,-0.35) {\small $u^*$};\node at (3.5,0.65) {\small $u^*$};

\foreach \x in {7.5,9.5} { 
\draw[red, line width=1,>-,>=latex] (\x,-0.5)  -- (\x,1.25);
}
\node at (4.25,-1) {\ldots};
\node at (8.5,-1) {\ldots};
\node at (1,-1) {\small $1$};
\node at (3,-1) {\small  $2$};
\node at (5.5,-1) {\small $j$};
\node at (7.5,-1) {\small  $j+1$}; 
\node at (9.5,-1) {\small  $L$};
\end{tikzpicture}
\end{center}
We have omitted $z$ in the sum over $\alpha$ in \eqref{Treduced} because the same argument used to obtain \eqref{TSvanish} yields $\mathcal{T}^{[\ldots j]}_z = 0$. 

To evaluate the subsystem transfer matrices, we note that \eqref{eq:LzL} generalises to
\be 
\mathcal{L}(u^*) \sigma_{\rm a}^\alpha  \mathcal{L}(u^*)   
= 
\frac{1}{\Delta_S^2}
\sum_{\beta = 0,\pm,z} \sigma_{\rm a}^\beta
 A_j^{\beta \alpha}
\,,
\qquad \alpha = 0,\pm \ ,
\label{eq:LsigmaL}
\ee 
where 
\be
\begin{gathered}
A_j^{0,0} =  1-\Delta_S^2 + \cosh\eta \cosh(2 \eta S_j^z)\ ,\qquad
A_j^{\pm,0} =  - 2 i \sinh\eta S_j^\mp \cosh(\eta S_j^z \mp \tfrac{\eta}{2})\ , \\
A_j^{0,\pm} = - \frac{i}{2} \sinh(2\eta) S_j^\pm \cosh(\eta S_j^z \pm \tfrac{\eta}{2})\ , \qquad\quad
A_j^{\pm,\pm} = \frac{\cosh\eta +\cosh(2\eta S_j^z)}{2}\ , \\
A_j^{\pm,\mp} = - \sinh^2 \eta (S_j^\mp)^2\ ,\qquad
A_j^{z,0} = \sinh^2\eta \Sigma_j^0   \,, \qquad A_j^{z,\pm} = \frac{\sinh^2\eta \cosh\eta}{2} \Sigma_j^\pm\ .
\end{gathered}
\ee
These imply the recursion relations
\be 
\mathcal{T}^{[\ldots j]}_\alpha =\frac{1}{\Delta_S^2} \sum_{\beta=0,\pm} \mathcal{T}^{[\ldots j-1]}_\beta A_j^{\beta \alpha}\ .
\ee 
for $\alpha=0,\pm$. Since the left boundary condition requires that $\mathcal{T}_{\alpha}^{[\ldots 0]} = \mathrm{tr}_{\rm a}(\sigma_{\rm a}^\alpha) = 2\delta_{\alpha0}$\;, the subsystem transfer matrices can be written as 
\begin{align}
\mathcal{T}^{[\ldots j]}_\alpha =  2\Delta_S^{-2j} \sum_{\alpha_1,\dots,\alpha_{j-1}\in\{0,\pm\}} 
A_1^{0\alpha_1}A_2^{\alpha_1\alpha_2}\dots A_j^{\alpha_{j-1}\alpha}\ .
\end{align}
We now define the ESZM operator as
\be
\Psi\equiv \frac{i}{2\cosh\eta \cosh\xi^-}{T^{(S,\frac12)}}'(u^*)=\sum_{j=1}^L\Psi_j.
\label{Psisum}
\ee
where each operator $\Psi_j$ acts non-trivially only on sites $1,\dots, j$ of the lattice. They are
\begin{align}
\label{Psinormalizedbis} 
\Psi_{j}&=\frac{\sinh^2\eta}{\Delta_S^{2j}}
\sum_{\alpha_1,\dots,\alpha_{j-1}\in\{0,\pm\}} A_1^{0\alpha_1}A_2^{\alpha_1\alpha_2}\dots A_{j-1}^{\alpha_{j-2}\alpha_{j-1}} \Sigma_j^{-\alpha_{j-1}}\ ,\quad 1\leq j<L\ ,\\ \nonumber
\Psi_L&=\frac{\sinh^2\eta}{\Delta_S^{2L}}
\sum_{\alpha_1,\dots,\alpha_{L-1}\in\{0,\pm\}} A_1^{0\alpha_1}A_2^{\alpha_1\alpha_2}\dots A_{L-1}^{\alpha_{L-2}\alpha_{L-1}} \left[\Sigma_L^{-\alpha_{L-1}}+\frac{\tanh\xi^-}{\sinh\eta\cosh \eta}A_L^{\alpha_{L-1}0}\right]\nn
&\equiv \overline{\Psi}_L+\tanh\xi^-B_{\rm R}\ .
\label{BRdef} 
\end{align}
The terms in the last line are defined so that $\overline{\Psi}_L$ is odd under spin flip, while  $B_{\rm R}$ is even. Indeed, this operation turns $A_j^{\alpha,\beta}$ into $A_{j}^{-\alpha,-\beta}$ and $\Sigma_j^{\alpha}$ into $-\Sigma_j^{-\alpha}$ for $\alpha,\beta \in \{0,\pm\}$. Therefore, all the $\Psi_j$ with $j<L$ are odd as well. One then can define an SZM, as we explain in section \ref{sec:ESZMSZM}. 
In the $S=\tfrac12$ case we recover the results of Refs \cite{fendley2016strong,FGVV}.

\subsubsection{ESZM as a matrix product operator and normalizability}
\label{sec:MPO}

Here and in the next subsection \ref{ssec:locality} we derive several key properties of arbitrary-$S$ ESZM defined by \eqref{Psinormalizedbis}. The statements here are for the operator norms, not for the operators themselves. We already indicated that because of the odd number of ground states for integer $S$, it is impossible for the associated SZM to square to the identity.  Here we show that the ESZM $\Psi$ remains normalizable throughout the gapped phase ($\eta \in \mathbb{R}$) even as $L\to\infty$. Moreover, we will show that it squares to the identity up to corrections exponentially small in the sense of the Hilbert-Schmidt norm
\be 
\lim_{L\to\infty}\big\| \Psi^2 - \mathds{1} \big\|^2 = 0 \,.
\label{eq:Psisquaresid}
\ee 
This property is a weaker version of $\lim_{L\to\infty}{\Psi^{(\frac12)}}^2= \mathds{1}$, but is consistent with e.g.\ the odd number of ground states for integer $S$.
Since $\Psi$ has a finite overlap with operators localized at the left boundary, this is a strong indication that it has good locality properties, an issue we address in more detail below. 

In order to prove \eqref{eq:Psisquaresid}, it is convenient to write $\Psi$ as a Matrix Product Operator (MPO), as follows directly from our transfer matrix construction in the previous section. The MPO form of $\Psi$ involves a 4-dimensional ancillary space $\mathfrak{A}$ whose basis states we label as $|0\rangle,|+\rangle,|-\rangle,|z\rangle$, where we emphasise that these kets are for vectors in $\mathfrak{A}$, not the physical Hilbert space of the model. We then define a linear operator $\mathcal{A}_j$ acting on  $\mathfrak{A}$, each of whose entries acts on the physical Hilbert space. In this basis it can be written as the matrix
\be
{\cal A}_j = \frac{1}{\Delta_S^2}\left(
\begin{array}{cccc}
{A_j^{0,0}} & {A_j^{0,+}} &{A_j^{0,-}} &{\Delta_S^2} \Sigma_j^0  \\
{A_j^{+,0}} &  {A_j^{+,+}} &{A_j^{+,-}}&{\Delta_S^2}  \Sigma_j^+ \\
{A_j^{-,0}} & {A_j^{-,+}}  &{A_j^{-,-}}&{\Delta_S^2}  \Sigma_j^{-} \\
0 & 0 & 0 & {\Delta_S^2}\mathds{1}
\end{array}
 \right)
\,. \ee
Then the ESZM and its conjugate can be written as
\be 
\Psi = \langle v_{\rm L}  | 
 {\cal A}_1  {\cal A}_2 \dots  {\cal A}_L
   | v_{\rm R}  \rangle  
\,,\qquad
 \Psi^\dagger = \langle v_{\rm L} | 
 {\cal A}_1  {\cal A}_2 \dots  {\cal A}_L
  | v_{\rm R}^* \rangle \,,
 \ee 
where utilise the vectors 
\be
\langle v_{\rm L} |  =\langle 0 | \,,
  \qquad |v_{\rm R}\rangle  = \frac{\sinh^2\eta}{\Delta_S^2} |z\rangle + \tanh\eta \tanh\xi^- |0\rangle
\ee
in $\mathfrak{A}$.
This expression shows that $\Psi$ is Hermitian whenever $\xi^- \in \mathbb{R}$. The MPO formulation provides us with a convenient expression for the Hilbert-Schmidt norm of $\Psi$ in terms of a trace
$\mathrm{Tr}$ over the physical space:
\begin{equation}
\begin{gathered}
\big\|\Psi\big\|^2\equiv \frac{1}{(2S+1)^L}\mathrm{Tr}(\Psi\Psi^\dagger) =
\big(\langle v_{\rm L} | \otimes\langle v_{\rm L} | \big) 
\,\mathbb{U}^L\,
\big(| v_{\rm R} \rangle\otimes | v_{\rm R}^* \rangle\big) \ ,\nn
\mathbb{U} = \frac{1}{2S+1} \mathrm{Tr}({\cal A}\otimes {\cal A}) \ .
\label{eq:HSnorm}
\end{gathered}
\end{equation}
Graphical representations of the trace and the matrix $\mathds{U}$ are
\tikzstyle{tensor}=[circle,draw=blue!50,fill=blue!20,thick]
\tikzstyle{vl}=[circle,draw=black!80,fill=black!80,thick]
\tikzstyle{vr}=[rectangle,draw=black!80,fill=black!80,thick]
\tikzstyle{vrs}=[rectangle,draw=black!80,fill=white,thick]
\begin{center}
\begin{tikzpicture}[inner sep=1mm]
 \node at (-1.1,0.5) { ${\rm Tr}(\Psi\Psi^\dagger)=$};
 \draw[blue,-] (0.2,0) -- (0.7,0);
    \draw[blue,-] (0.2,1) -- (0.7,1);
    \draw[blue,-] (8.2,0) -- (8.7,0);
    \draw[blue,-] (8.2,1) -- (8.7,1);
    \node[vl] at (0.2,0){};
    \node[vl] at (0.2,1){};
    \node[vr] at (8.7,0){};
    \node[vrs] at (8.7,1){};
     \foreach \i in {1,...,8} {
        \draw[red] (\i+0.1,0.5) ellipse (0.1cm and 1.5cm);
        \node[tensor] (\i) at (\i, 0) {${\cal A}$};
        \node (\i spin) at (\i, 1) {};
        \draw[red,-] (\i) -- (\i spin);
    };
    \foreach \i in {1,...,7} {
        \pgfmathtruncatemacro{\iplusone}{\i + 1};
        \draw[blue,-] (\i) -- (\iplusone);
    };
    \foreach \i in {1,...,8} {
        \node[tensor] (\i) at (\i, 1) {${\cal A}$};
    };
    \foreach \i in {1,...,7} {
        \pgfmathtruncatemacro{\iplusone}{\i + 1};
        \draw[blue,-] (\i) -- (\iplusone);
    };
\end{tikzpicture}
\label{fig:diagram2}
\qquad\quad
  \begin{tikzpicture}
\node at (-0.3,0.75) { $\mathds{U}=$};
  \draw[red,-] (1,0.25) -- (1,1.75);
  \draw[blue,-] (0.25,1.5) -- (0.75,1.5);
  \draw[blue,-] (1.25,1.5) -- (1.75,1.5);
  \draw[blue,-] (0.25,0) -- (0.75,0);
  \draw[blue,-] (1.25,0) -- (1.75,0);
   \draw[red] (1.1,0.75) ellipse (0.1cm and 1.5cm);
    \node[tensor] at (1,0){$\cal A$} ;
  \node[tensor] at (1,1.5){$\cal A$} ;
  \end{tikzpicture}
\end{center}
%
We have verified by direct calculation that for any $S$ and for $\eta\in \mathbb{R}$ the matrix $\mathbb{U}$ has a unique eigenvalue equal to 1 and all other eigenvalues are strictly smaller in absolute value.
Its left and right eigenvectors in $\mathfrak{A}\otimes \mathfrak{A}$ are: 
\begin{align}
| \mathrm{R} \rangle &= \left( \tfrac{\Delta_S}{\sinh\eta}\right)^4 
\left(|0 \rangle \otimes | 0 \rangle - \tfrac{2}{\cosh^2 \eta}   (|+\rangle \otimes | - \rangle\ +\ |- \rangle \otimes | + \rangle )\right)\  +\ |z\rangle \otimes | z \rangle\ ,\nn
\langle \mathrm{L} | &= \langle z | \otimes \langle z |\ .
\end{align} 
These are normalized so that $\langle  \mathrm{L} |  \mathrm{R} \rangle =1$, and 
\be 
\langle  \mathrm{L} |   v_\mathrm{R} \otimes v_\mathrm{R} \rangle
\langle     v_\mathrm{L} \otimes v_\mathrm{L} |  \mathrm{R} \rangle=1.
\ee 
We conclude that for asymptotically large $L$
\be
\big\|\Psi\big\|^2= 1+{\cal O}(e^{-\gamma L})\ ,
\ee
where the exponentially small corrections are determined by the largest subleading eigenvalue of $\mathbb{U}$.

We are now in a position to prove \eqref{eq:Psisquaresid}. We start by writing out the norm
\begin{align} 
\big\|\Psi^2 -  \mathds{1} \big\|^2 
=
\tfrac{1}{(2S+1)^L} \mathrm{Tr}\big(\big(\Psi^2\big)^\dagger\Psi^2\big)
- 
\tfrac{1}{(2S+1)^L} \Big[\mathrm{Tr}\big(\big(\Psi^2)^\dagger)+\mathrm{Tr}\big(\Psi^2\big)\Big]+1
\ .
\label{eq:HSnorm2}
\end{align}
The traces quadratic in $\Psi$ can be evaluated in the same way as \eqref{eq:HSnorm}: they involve the same ancillary transfer operator $\mathbb{U}$ and only differ in the choice of the right boundary vector. As a result, they both tend (with exponential accuracy) to $1$ in the thermodynamic limit. This leaves us with
\be 
\tfrac{1}{(2S+1)^L}\mathrm{Tr}\Big(\big(\Psi^\dagger\big)^2\Psi^2\Big) =
\langle w_{\rm L}  |  
\mathbb{V}^L
| w_{\rm R}\rangle \,,
\ee 
where 
\[
\mathbb{V} = \tfrac{1}{2S+1} \mathrm{Tr}({\cal A}\otimes {\cal A} \otimes {\cal A}\otimes {\cal A}) \ ,\quad
\langle w_{\mathrm{L}} | = \langle v_{\mathrm{L}} |^{\otimes 4}\ ,\quad
|w_{\rm R} \rangle  = (|v_{\rm R}\rangle^*)^{\otimes 2} \otimes(|v_{\rm R}\rangle)^{\otimes 2}\ .
\]
For $\eta \in \mathbb{R}$ the matrix $\mathbb{V}$ has one eigenvalue equal to $1$ 
and all others have absolute values strictly smaller than 1. The left eigenvector associated with eigenvalue $1$ is $\langle  \mathtt{L} | = \langle z | \otimes \langle z |\otimes \langle z |\otimes \langle z |$. The right eigenvector $|\mathtt{R}\rangle$ may also be written out explicitly, but we only need the following property
\be 
\frac{
\langle  \mathtt{L} |  w_\mathrm{R} \rangle
\langle    w_\mathrm{L} | \mathtt{R} \rangle}{\langle  \mathtt{L} | \mathtt{R} \rangle}
= 1.
\ee 
Putting everything together we conclude that the right-hand side of \eqref{eq:HSnorm2} vanishes, up to exponential corrections determined by the subleading eigenvalues of $\mathbb{U}$ and $\mathbb{V}$. We have thus shown that the norm of the ESZM for general $S$ acts similarly to the way the operator itself does for $S$\,=\,$\tfrac12$.

\subsubsection{Spatial locality properties of the ESZM operator}
\label{ssec:locality}
\begin{figure}[ht]
\centering
\includegraphics[scale=0.75]{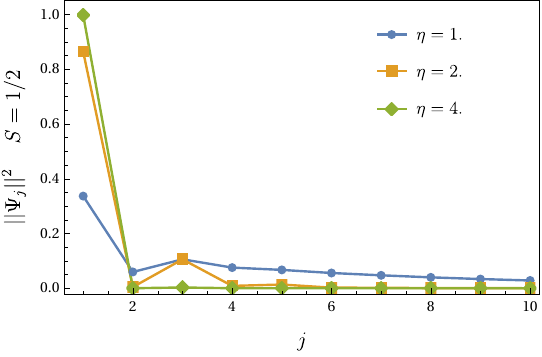}
~~
\includegraphics[scale=0.75]{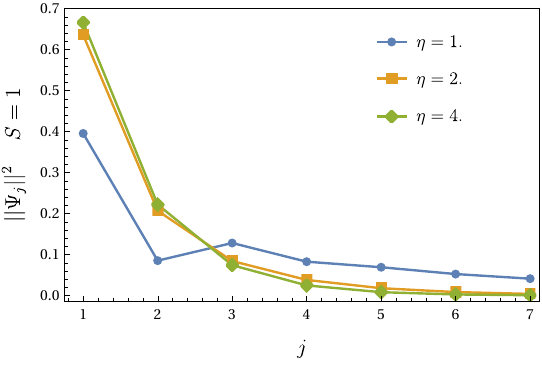}
\caption{\small Hibert-Schmidt norm of $\Psi_j$ from \eqref{Psinormalizedbis}, for $S=\tfrac12$ (left panel) and $S=1$ (right panel).}
 \label{fig:psijnorm}
 \end{figure}

We now investigate the spatial locality properties of our ESZM operator.  As we have shown in \eqref{Psisum}, $\Psi$ can be written as
as sum over operators $\Psi_j$ that acts trivially to the right of $j$. 
The Hilbert-Schmidt norm of $\Psi_j$ can be computed along the same lines as in the computation of the norm of $\Psi$, utilising the same ancillary space $\mathfrak{A}\otimes \mathfrak{A}$ and transfer operator $\mathbb{U}$. As apparent from their explicit expression \eqref{Psinormalizedbis}, the $z$ channel does not appear in the rightmost part of $\Psi_j$, so the right-boundary vectors needed are the form $| \alpha \rangle \otimes | \beta \rangle$ with $\alpha,\beta \in \{0,\pm\}$. These all project the ancillary space orthogonally to the leading left-boundary eigenvector $\langle \mathtt{L} |$, therefore $\big\|\Psi_j\big\|^2$ decays exponentially with $j$, with an amplitude determined by the first subleading eigenvalue of $\mathbb{U}$.
A numerical demonstration of this exponential falloff is presented in Fig. \ref{fig:psijnorm} for $S=\tfrac12$ and $S=1$. 

This exponential falloff is weaker for $S\geq 1$ than the $S=\tfrac12$ SZM operators previously constructed. This behaviour is apparent in the limit $\eta\to\infty$, where
\be
\lim_{\eta\to\infty}\Psi_j=\begin{cases}
\sigma^z_1 \delta_{j,1}& \text{if } S=\tfrac12\ ,\\
E_1\dots E_{j-1}\Sigma_j  &\text{if } S\geq 1\ .
\end{cases}
\label{psij}
\ee
with 
\begin{align}
E_j&=\mathrm{diag}(0,-1,\ldots -1, 0)\ ,\quad
\Sigma_j=\mathrm{diag}(1,0,\ldots 0, -1)
\label{psiinfiniteeta}
\end{align}
acting on the physical states. In this limit, the $S=\tfrac12$ SZM operator becomes completely localized at the left boundary, while for $S\geq 1$ it acts non-trivially far away from the left boundary. While the latter action for $j$ large is non-trivial on an exponentially large number of states, their number is still negligible compared to the full Hilbert space. Hence this behaviour is still consistent with the vanishing norms derived above. 

\begin{figure}[ht]
\centering
\includegraphics[scale=0.75]{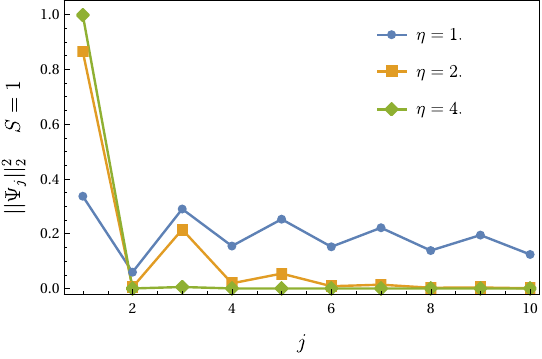}
~~
\includegraphics[scale=0.75]{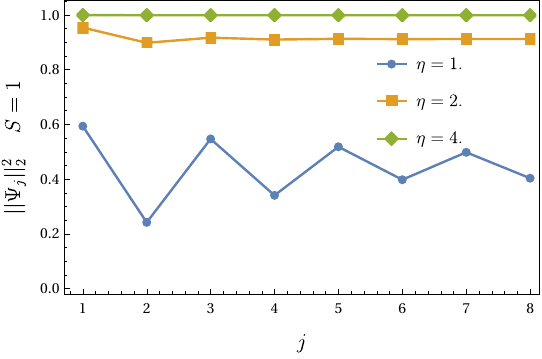}
\caption{\small Spectral norm $\big\|\Psi_j\big\|_2^2$, as defined in \eqref{eq:norm2}, for $S=\tfrac12$ (left panel) and $S=1$ (right panel).}
 \label{fig:psijinfinitynorm}
 \end{figure}

This lack of exact localisation for $S>\tfrac12$ 
can be quantified further by considering a different operator norm, e.g. the spectral norm
\be
\big\|\Psi_j\big\|_2=\sqrt{\lambda_{\rm max}(\Psi^\dagger_j\Psi_j)}\ .
\label{eq:norm2}
\ee
Our numerical checks indicate that for $S=\tfrac12$ this norm decays exponentially in $j$, while for $S\geq 1$ it does not, as illustrated in Fig.~\ref{fig:psijinfinitynorm}.
 
Nevertheless, the ESZM leads to clear signatures in edge autocorrelation functions $C_0^{\cal O}(t)$ as shown for spin 1 in Fig.~\ref{fig:autocorr}. In particular, these autocorrelators converge to finite plateaux values, which in some cases are determined by the overlap of the operator ${\cal O}$ with the ESZM, \emph{cf.} Ref.~\cite{vernier2024strong,gehrmann2025exact}.
To compute these values, we consider infinite-temperature autocorrelation functions \fr{autocorr}
for local operators ${\cal O}(0)$ that act non-trivially only very close to the left boundary. 
To see the effects of an orthonormal set $\{\Psi^{(\alpha)}\}$ of Hermitian ESZM operators localized around the left boundary on such correlators we decompose ${\cal O}(0)$ as
\be
{\cal O}(0)=\sum_\alpha c^{\cal O}_\alpha{\Psi}^{(\alpha)}+{\cal O}'\ ,\quad {\rm Tr}\big({\Psi}^{(\alpha)}{\cal O}'\big)=0\ ,\quad
c^{\cal O}_\alpha=\frac{1}{(2S+1)^L} {\rm Tr}({\Psi}^{(\alpha)} {\cal O})\,.
\ee
where ${\cal O}'$ is by construction localized around the left boundary (possibly with a weaker form locality, as discussed above). Using that $\{\Psi^{(\alpha)}\}$ form an orthonormal set with respect to the Hilbert-Schmidt norm we then have
\be
C^{\cal O}_0(t)=\sum_\alpha|c^{\cal O}_\alpha|^2+C^{{\cal O}'}_0(t)\ .
\label{argument}
\ee
As by construction there are no other conserved quantities localized around the left boundary $C^{{\cal O}'}(t)$ is expected to decay in time to a value that vanishes as $L\to\infty$, which in turn implies that $C^{\cal O}(t)$ decays to a finite value $\sum_\alpha|c^{\cal O}_\alpha|^2$ set by the overlaps of ${\cal O}(0)$ with $\Psi^{(\alpha)}$. We now apply these considerations to the numerical results presented in section \ref{sec:auto}. In section \ref{ssec:TShalf} we have constructed a single ESZM operator $\Psi$. We then conclude that
\be
\lim_{t\to\infty}\lim_{L\to\infty}C^{\cal O}_0(t)\geq \Big|\frac{1}{(2S+1)^L}{\rm Tr}\big(\Psi{\cal O}\big)\Big|^2\equiv |c^{\cal O}_1|^2.
\label{inequality}
\ee
\begin{itemize}
\item{}$S=\tfrac12$: Here our numerical results are compatible with an equality in \fr{inequality}. An example is shown by the dashed red line in the left panel of Fig.~\ref{fig:autoxxz_hl}.
\item{}$S=1$: Here our numerical results indicate that there must be additional ESZM operators. For boundary conditions \fr{bpot} the overlaps of our ESZM operator with $s^z_1$ and ${\cal O}_2=(1-(s^z_1)^{2})s^z_2$ are given by
\be
c^{s^z_1}_1=\tfrac{2}{3}\Big[1-\tfrac{1}{(1-2\cosh\eta)^2}\Big]\ ,\quad
c^{{\cal O}_2}_1=\tfrac{2}{9}\Big[-1-\tfrac{2}{(1-2\cosh\eta)^4}+\tfrac{3}{(1-2\cosh\eta)^2}\Big].
\ee
For $\eta=1$ these give $|c^{s^z_1}_1|^2=0.263665$ and $|c^{{\cal O}_2}_1|^2=0.00855693$, which are much smaller than the corresponding plateau values seen in Fig.~\ref{fig:autocorr}. The fact that there must be additional ESZM operators is most easily seen in the strong coupling limit $\eta\to\infty$. Here the Hamiltonian becomes classical, so that $s^z_1$ becomes a conserved quantity. However, the ESZM operator \fr{Psisum} only has an overlap of $2/3$ in this limit, and cannot by itself account for the conservation of $s^z_1$.
\end{itemize}

\subsubsection{ESZM for the \sfix{$\mathbb{Z}_2$-invariant boundary condition $\xi^+=0$}}
\label{sec:EZSMxipluszero}

For integer $S$, another choice of left boundary condition invariant under spin flip is $\xi^+=0$. The numerical results presented on Figure \ref{fig:autocorr_x} for the autocorrelation functions in the $S=1$ chain suggest that an ESZM should exist in that case too.  
In that case, the left-reflection matrix becomes $K_{\rm a}^+(u)= \sinh(u+\eta)\sigma_{\rm a}^z$, and indeed, we check that all the technical steps detailed above for $\xi^+=i\pi/2$ can be reproduced after having exchanged the roles of $\sigma_{\rm a}^0$ and $\sigma_{\rm a}^z$. In particular, the cornerstone equation \eqref{eq:LzL} holds after making that replacement and setting $u^\ast=0$. 
The construction results in another operator $\Psi'$ playing the role of an ESZM for the boundary condition $\xi^+=0$ (although $\Psi'$ can be defined for any $S$, it plays the role of an ESZM for integer $S$ only, as for $S\in \mathbb{Z}_+ +1/2$ the Hamiltonian with boundary condition $\xi^+=0$ is not properly defined). We will not give its complete expression here, but simply mention that its leading order contribution in a large $\eta$ expansion is the same as that of $\Psi$, namely given by eq.~\eqref{psij}.

\subsection{ESZM vs SZM}
\label{sec:ESZMSZM}
In the spin-$\tfrac12$ case there is a simple relation between the ESZM and SZM operators, \emph{cf.} the discussion surrounding eqn \fr{ESZMvsSZM}. A natural question is whether it is possible to construct SZM operators for higher spins as well. We recall that for $S=1/2$ the SZM operator arises when both boundaries are equivalent with vanishing boundary fields $h_1=h_L=0$, and can be obtained from the ESZM operator by subtracting the contribution arising from the far boundary, which diverges when $h_L\to 0$.
We observe that for spin $S$ the ESZM $\Psi$ has a diverging contribution in the limit $\xi^-\to i\pi/2$. In this limit the two boundaries become equivalent with boundary potentials
\be
b^{(1)}_1=\big(1+J_z-\sqrt{2(1+J_z)}\big)(S^z_1)^2\ ,\qquad
b^{(1)}_L=\big(1+J_z-\sqrt{2(1+J_z)}\big)(S^z_L)^2.
\ee

We deal with this situation in analogy to the $S=\tfrac12$ case by defining a SZM operator
\be
\overline{\Psi}\equiv\Psi-\tanh\xi^-B_{\rm R}\ ,
\label{SZM_S1}
\ee
where $B_{\rm R}$ is defined in \fr{BRdef}. We can see that this fulfills the requirement \eqref{expsmall} for a SZM operator for $S$\,=\,1 as follows (the argument for general $S$ is analogous). Given that $[H^{(1)}(\xi^-),\Psi]=0$ for arbitrary $\xi^-\neq \tfrac{i\pi}2$ we must have $[H^{(1)}\big(\tfrac{i\pi}2\big),B_{\rm R}]=0$ in order for the most divergent contribution to the commutator to vanish. The remaining terms in the limit $\xi^-\to i\frac{\pi}{2}$ of the commutator $[H^{(1)}(\xi^-),\Psi]$ then give
\be
\big[H^{(1)}\big(\tfrac{i\pi}2\big),\overline{\Psi}\big]=\frac{2\sinh2\eta}{1+\cosh\eta}[B_{\rm R},s^z_L]\ .
\ee
Using the MPO representation one can show that the Hilbert-Schmidt norm of $B_{\rm R}$ and therefore of the right-hand-side is exponentially small in system size, which in turn yields
\be
\big\|[H^{(1)}\big(\tfrac{i\pi}2\big),\overline{\Psi}]\big\|={\cal O}(e^{-\gamma L})\ .
\label{SZMS1}
\ee
In contrast to the $S=\tfrac12$ case, we do not expect the commutator to be exponentially small in other operator norms such as the spectral norm \fr{eq:norm2}. Moreover, as with $\Psi$, the SZM $\overline{\Psi}$ is localised near the edge only in the weak sense. Likewise, we know from the $\eta\to\infty$ limit \fr{psij} that the norm of $\overline{\Psi}^2$ only satisfies the weaker condition
\be
\big\|\overline{\Psi}^2-\mathds{1}\big\|={\cal O}(e^{-\gamma' L})\,.
\ee
For all $\eta$ real and positive this result follows from the triangular identity, combining the definition \eqref{SZM_S1} with \eqref{eq:Psisquaresid} and the exponential decrease of the norm $\| B_{\rm R} \|$. This SZM does anticommute with the spin-flip operator, as follows from the discussion after \eqref{BRdef}. 

The existence of a SZM operator in the sense of eqn \fr{SZMS1} has observable consequences for infinite-temperature autocorrelation functions. As shown in Fig.~\ref{fig:autocorrS1_SZM}, $C^{S^z_1}_0(t)$ approaches a plateau value whose duration in time grows with system size, and only at later times exhibits a very slow decay like the one seen for non-integrable boundary conditions, \emph{cf.} Fig.~\ref{fig:S1nonintBCs}.

\subsection{ESZM operators from transfer matrices \sfix{$T^{(S,S'>\frac{1}{2})}(u)$}}
\label{sec:higherauxspin}

The ESZM described above is constructed from the transfer matrix $T^{(S,S')}(u)$, with auxiliary spin $S'=\frac{1}{2}$ and physical spin $S$. A natural question is whether one can obtain other ESZMs from other values of $S'$. Our work indicates that only for some values of $(S,S')$ it is possible. 
For the simplest case $(S,S')$\,=\,($\tfrac12$,1) the answer appears to be negative, as we summarize in Appendix~\ref{app:ShalfSprime}. 

However, there are good signs that at least for physical spin $S=\tfrac32$, a distinct ESZM exists. 
One indication is obtained by considering a strong-coupling expansion around the $\eta\to\infty$ limit, following the method employed in the $S=\tfrac12$ chain in Refs. \cite{fendley2016strong,kemp2017long}. 
We described the ground states and low-energy states at large $\eta$ in section \ref{sec:spin32}. Since there an even number of the former, it is possible for an SZM in the traditional sense to occur. 

Guided by the $S=\tfrac12$ case \cite{fendley2016strong}, we develop a strong-coupling expansion of a putative (E)SZM operator
\be
\Sigma=\Sigma_0+e^{-\eta}\Sigma_1+e^{-2\eta}\Sigma_2+e^{-3\eta}\Sigma_3+\dots\ .
\ee
The idea is now to use the strong-coupling expansion of the Hamiltonian
\be
H^{(\frac32)}=e^{\eta} H_0+H_1+e^{-\eta} H_2+\dots
\ee
together with the condition $[\Sigma,H]=0$ to derive the operators $\Sigma_j$ order-by-order. At leading order we need $[\Sigma_0,H_0]=0$. We could pick $\Sigma_0=S^z_1$, but find there is a simpler operator that works better. Namely, we instead take
\be
\Sigma_0=P_1^{1}+P_1^{2}-P_1^{3}-P_1^{4}
\label{sigma0}
\ee
using the conventions of section \ref{sec:spin32}. This operator anticommutes with spin-flip like  $S_1^z$, but has only two distinct eigenvalues.\
The leading term $H_0$ is in \eqref{H032}, and
\begin{align} H_1=\frac{1}{2}\sum_j\left[E_j^{12}E_{j+1}^{43}
+E_j^{23}E_{j+1}^{32} +E_j^{34}E_{j+1}^{21} +{\rm h.c.}\right]\ ,
\end{align}
where $E_j^{ab}=|a_j\rangle\,\langle b_j|$ so that $P_j^a=E^{aa}_j$. The next step is to find a $\Sigma_1$ that solves 
\be
\big[\Sigma_1,H_0\big]=
E_1^{32}E_2^{23}-E_1^{23}E_2^{32}=\big[H_1,\Sigma_0\big]\ .
\ee
We find 
\be
\Sigma_1=-4\big(E_1^{23}E_2^{32}+E_1^{32}E_2^{23}\big)\big(P_3^{1}+P_3^{2}-P_3^{3}-P_3^{4}\big)
\label{sigma1}
\ee
works. 
This expression is very similar to the one encountered in the strong-coupling expansion of the SZM in the spin-$\tfrac12$ XXZ chain \cite{fendley2016strong}, but we do not expect this to persist for the higher orders $\Sigma_{j\geq 2}$.
The strong coupling expansion suggests the existence of a ``conventional'' ESZM, having the same spatial structure as in the spin-$\tfrac12$ case. The structure of this operator is very different from the ESZM $\Psi$ constructed from $T^{(S,\frac12)}$, where the leading term in the strong-coupling expansion \fr{psiinfiniteeta} involves operators acting on all sites of the lattice. 

Motivated by this result, we investigate whether we can obtain another (E)SZM for $S$\,=\,$\tfrac32$ using our transfer-matrix construction with higher auxiliary spin. Such transfer matrices can constructed utilising a set of recurrence relations known as the T-system, which provides an expression of $T^{(S,S')}$ in terms of products of $T^{(S,\frac12)}$, or, more conveniently, of the rescaled transfer matrix 
\be 
t^{(S)}(u) = \frac{T^{(S,\frac12)}(u)}{\omega_1(u)}  \,, \qquad \omega_1(u)=\frac{\sinh(2u+2\eta)\sinh(u+\xi^+)\sinh(u+\xi^-)}{\sinh(2u+\eta)}
\ee 
(see Appendix~\ref{app:Tsystem} for details). 
Since $\omega_1(i\pi/2)=0$ for $\xi^- = i\pi/2$, the previously derived ESZM, which is proportional to ${T^{(S,\frac12)}}'\big(\tfrac{i\pi}2\big)$, is simply proportional to $t\big(\tfrac{i\pi}2\big)$ with no derivative. 

We conjecture that for $S=\tfrac32$ another ESZM with good locality properties is obtained  by
\be 
\Psi_{\tfrac32} = T^{(\tfrac32,\tfrac32)}\big(\tfrac{i\pi}2\big) \ ,
\label{eq:Psi32def}
\ee 
where $T^{(\tfrac32,\tfrac32)}(u)$ is constructed from the T-system, and defined as 
\be 
\begin{split}
T^{(\tfrac{3}{2},\tfrac{3}{2})}(u) = 
t^{(\frac{3}{2})}(u-\eta) t^{(\frac{3}{2})}(u) t^{(\frac{3}{2})}(u+\eta) -f(u) t^{(\frac{3}{2})}(u-\eta) -f(u-\eta) t^{(\frac{3}{2})}(u+\eta)\ .
\end{split}
\ee 
In the above equation, the function $f$ is that defined in Appendix for generic physical spin $S$ and boundary parameters $\xi^\pm$ as 
\begin{align}
f(u) = \frac{\sinh(2u)\sinh(u-\xi^++\eta)\sinh(u-\xi^-+\eta)}{\sinh(2u+2\eta)\sinh(u+\xi^+)\sinh(u+\xi^-)} \left( \frac{\sinh(u+(\tfrac{1}{2}-S)\eta)}{\sinh(u+(\tfrac{1}{2}+S)\eta)} \right)^{2L} \,
\end{align}
(here those parameters must be fixed to $S=\tfrac32$, $\xi^- = i\pi/2$).
The operator $\Psi_{3/2}$ by construction commutes with $H^{(3/2)}$ \fr{H_S} and the transfer matrices $T^{(\tfrac32,S')}(u)$. Its large-$\eta$ expansion reads
\begin{align}
\Psi_{\tfrac32} &= \Sigma_0+e^{-\eta}\Sigma_1+ O(e^{-2\eta})\ ,
\end{align}
where $\Sigma_{0,1}$ are given by \fr{sigma0} and \fr{sigma1} respectively, recovering the result of the perturbative construction discussed above. We stress that the perturbative expansion does not distinguish between SZM and ESZM until the expansion reaches the far edge.
It would be interesting to carry out a full, explicit evaluation of \fr{eq:Psi32def}, but this is beyond the scope of our work.

\section{Strong Zero Mode and boundary bound states in the open spin-\sfix{$\tfrac12$} XXZ chain}
\label{sec:BBS}
A well-known effect of boundaries with non-trivial boundary phase-shifts is the possibility of the formation of spatially localized bound states in the ground state sector. In integrable models such bound states are associated with particular root configurations in the solutions of the Bethe equations. These are known as \emph{boundary strings}, and their effects on the low-energy physics have been extensively studied in the literature, see e.g. Refs\cite{kapustin1996surface,asakawa1996finite,essler1997x,bedurftig1997spectrum,deguchi1998gapless,Grijalva2019open,pasnoori2023boundary,kattel2024kondo}.

Ever since the construction of SZM operators in interacting integrable models \cite{fendley2016strong }, an obvious question has been how such boundary strings relate to SZM or ESZM operators. Given that we have obtained a formulation of the ESZM in terms of the transfer matrix we can finally answer this question. For technical reasons that become clear shortly we restrict our analysis to the spin-$\tfrac12$ case. Here boundary strings in the ground state have been analyzed comprehensively in the recent work \cite{Grijalva2019open}. 
\subsection{Bethe ansatz equations and boundary strings}
Eigenstates of the spin-$\tfrac12$ XXZ chain Hamiltonian (\ref{Hopenper},\ref{hhalf}) in the sector with $N$ down spins are
parametrized by $N$ rapidity variables $\alpha_j$, that are subject to the Bethe ansatz equations (BAE)
\cite{sklyanin1988boundary}
\be
\left[\frac{\sin\big(\alpha_j+i\frac{\eta}{2}\big)}
  {\sin\big(\alpha_j-i\frac{\eta}{2}\big)}\right]^{2L}
\prod_{\sigma=\pm}
\frac{\sin\big(\alpha_j-i\frac{\eta}{2}-i\txi_\sigma\big)}
{\sin\big(\alpha_j+i\frac{\eta}{2}+i\txi_\sigma\big)}
 =\prod_{k\neq j}\frac{\sin\big(\alpha_j-\alpha_k+i\eta\big)}
     {\sin\big(\alpha_j-\alpha_k-i\eta\big)}
\frac{\sin\big(\alpha_j+\alpha_k+i\eta\big)}
{\sin\big(\alpha_j+\alpha_k-i\eta\big)}     \ ,
\label{BAE}
\ee
where the parameters $\txi^\pm$ are given by the boundary fields (we follow the conventions of Ref.~\cite{Grijalva2019open} to facilitate comparisons with their analysis -- 
the relation to \fr{boundaryfieldshalf} is $\xi^\pm=\txi^\mp$)
\be
b^{(\frac{1}{2})}_1=-\sinh\eta\ \coth\txi^- \sigma^z_1\ ,\quad
b^{(\frac{1}{2})}_L=-\sinh\eta\ \coth\txi^+ \sigma^z_L\ .
\label{xipm}
\ee
We now impose
\be
\txi^-=\frac{i\pi}{2}\ ,\quad h_L\neq 0,\sinh\eta\coth(2\eta)\ .
\label{restriction}
\ee
The first two restrictions are required for the ESZM operator to exist, while the third simplifies the analysis in section~\ref{sec:GSS}. The BAE are unchanged under the replacements \emph{for an individual root}
\begin{align}
  \alpha_j&\rightarrow -\alpha_j\ ,\nn
  \alpha_j&\rightarrow \alpha_j+\pi\ .
\end{align}
We therefore identify solutions that are connected by such transformations.

It has been shown in Ref \cite{kapustin1996surface} that for a given solution  $\{\wta_1,\dots,\wta_N\}$ of \fr{BAE} some of the roots can be associated with singularities in the phase factors associated with the boundaries. For large $L$ these take the form of so-called \emph{boundary n-strings}
\be
\alpha_{n,j}^\sigma=-i\big(\tfrac{\eta}{2}+\txi_\sigma\big)-i(j-1)\eta\ ,\quad j=1,\dots,n.
\label{boundarystrings}
\ee
A solution of the BAE that contains a boundary $n$-string corresponds to a many-particle wave function that features a boundary bound state of magnons.

Given a solution $\{\wta_j\}$ of the Bethe equations for physical spin $S$\,=\,$\tfrac12$, the transfer-matrix form \eqref{Psisum} gives the associated eigenvalue of the ESZM operator $\Psi$ as
\be
\Psi(\{\wta_j\})=1-
\frac{\cosh(\eta+\txi^+)}{\cosh\eta\cosh\txi^+(\cosh\eta)^{2L}}
\prod_{k=1}^N
\frac{\cosh(3\eta)+\cos(2\wta_k)}{\cosh\eta+\cos(2\wta_k)}\ .
\label{ZMEV}
\ee
In the following we provide evidence in support of the following conjecture: 
\begin{itemize}
\item{} If the solution $\{\wta_j\}$ does not contain any boundary string associated with $\txi^-$ we have for large $L$
\be
\Psi(\{\wta_j\})=1+o(L^0)\ .
\label{psip1}
\ee
\item{} If the solution $\{\wta_j\}$ contains a boundary string associated with $\txi^-$ we have 
\be
\Psi(\{\wta_j\})=-1+o(L^0)\ .
\label{psim1}
\ee
\end{itemize}
Our conjecture is supported by a numerical analysis of the BAE for small systems sizes and numbers of roots up to $L=12$, $N=6$. Some results of these computations are presented in Appendix~\ref{app:numerics}. The numerics has no restrictions on the energy eigenvalues and low-lying excitations as well has very high energy states.

\subsection{Ground-state sector}
\label{sec:GSS}
Analytic results in the limit of large $L$ can be obtained in the \emph{ground state sector}, which refers to excited states with energies
\be
E=E_{\rm GS}+{\cal O}(L^0)\ ,
\ee
where $E_{\rm GS}$ denotes the ground state energy. This sector consists of states that involve fixed but arbitrary numbers of elementary excitations in the large-$L$ limit, but does not extend to eigenstates at finite energy densities relative to the ground state.

The structure of solutions of the BAE in the ground state sector has been determined in Refs~\cite{babelon1983analysis,kapustin1996surface}. In the following we focus on states with
\begin{itemize}
\item{} $M_r$ real roots $\alpha_j\in\mathbb{R}$, $1\leq j\leq M_r$, where $M_r$ scales with $L$;
\item{} $M_c$ roots ${\cal Z}=\{\lambda_j\}$ with non-zero imaginary part, where $M_c$ is fixed when $L$ increases; the set ${\mathcal{Z}}$ may contain at most two boundary roots $\alpha_{\rm BR}^\pm=-i\big(\frac{\eta}{2}+\txi_\pm+\epsilon_\pm\big)$. We do not allow longer "boundary strings" to keep the calculations simple.
\end{itemize}
The total number of roots is $N\leq L/2$ and the solutions are assumed to be compatible with the value of the boundary parameters, \emph{cf}. Ref.~\cite{Grijalva2019open} for a detailed discussion. By taking the logarithm the Bethe equations \fr{BAE} for real roots can be cast in the form
\begin{align}
\hat{\txi}(\alpha_j|\{\wta\})
&=\phi(\alpha_j,\frac{\eta}{2})-\frac{1}{2L}\sum_{\sigma=\pm}\phi(\alpha_j,\frac{\eta}{2}+\txi_\sigma)-\frac{1}{2L}\sum_{k\neq j}^{M_r}\Phi(\alpha_j-\alpha_k,\eta)\nn
&\qquad-\frac{1}{2L}\sum_{k=1}^{M_c}\Phi(\alpha_j-\lambda_k,\eta)=\frac{\pi I_j}{L}\ ,
\end{align}
where $1\leq I_j\leq M-1$ are integers and $M$ depends on $\txi_\pm$ \cite{Grijalva2019open} and
\be
\Phi(\alpha,\mu)=\phi(\alpha-\mu,\eta)+\phi(\alpha+\mu,\eta)\ ,\quad
\phi(\alpha,\zeta)=i\ln\left[\frac{\sin(i\zeta+\alpha)}{\sin(i\zeta-\alpha)}\right]\ .
\ee 
For a given solution $(\{\wta\})$ some of the allowed values of $I_j$ for real roots will not be taken. Denoting these integers by $h_1,\dots,h_{n_h}$, we define corresponding hole rapidities by
\be
\hat{\txi}(\check{\lambda}_{h_k}|\{\wta\})=\frac{\pi h_k}{L}\ ,\quad k=1,\dots, n_h.
\label{holes}
\ee
Using the structure of the solution $\{\wta\}$ discussed above we can rewrite \fr{ZMEV} in the form
\begin{align}
1-\Psi(\{\wta\})&\equiv e^{\chi(\{\wta\})}\ ,\nn
\chi(\{\wta\})&=
\ln\Big[\frac{\cosh(\eta+\txi^+)}{\cosh\eta\cosh\txi^+}\Big]
-2L\ln(\cosh\eta)-\sum_{k=1}^{M_r}\psi_-(\alpha_k)
-\sum_{k=1}^{M_c}\psi_-(\lambda_k)\ ,
\label{SZMEV}
\end{align}
where we have defined
\be
\psi_\sigma(\alpha)=\ln\Big[\frac{\cos(2\alpha)-\cosh(\eta-2\txi_\sigma)}
  {\cos(2\alpha)-\cosh(3\eta+2\txi_\sigma)}  \Big]\ .
\label{psisigma}
\ee
For large $L$ we now turn the sum over real roots into integrals, \emph{cf.} Appendix~\ref{app:sumtoint}, which gives
\begin{align}
\chi(\{\widetilde\alpha\})=&-2L\ln(\cosh\eta)-L\int_{-\frac{\pi}{2}}^{\frac{\pi}{2}}
\frac{d\alpha}{2\pi}\psi_-(\alpha)
        [\rho_0(\alpha)+\frac{1}{L}\rho_1(\alpha)]\nn
&- \ln\left[\frac{\sinh(3\eta)}{\sinh\eta}\right]
+\ln\Big[\frac{\cosh(\eta+\txi^+)}{\cosh\eta\cosh\txi^+}\Big]+\sum_{j=1}^{n_h}\psi_-(\check{\lambda}_{h_j})
-\sum_{k\in\mathcal{Z}}\psi_-({\lambda}_{k}).
\label{zeromodeEV}
\end{align}
The densities $\rho_{0,1}(\alpha)$ solutions to the linear integral equations \fr{rho01}.
The right hand side of \fr{zeromodeEV} has contributions that are proportional to $L$ as well as contributions that scale as $L^0$. Crucially, there can be a contribution to the extensive part on the r.h.s. if one of the complex roots $\lambda_k$ is a boundary root. The boundary roots are located at
\be
\alpha_{\rm BR}^\pm=-i\big(\frac{\eta}{2}+\txi_\pm+\epsilon_\pm\big)\ ,
\label{abr}
\ee
where $|\epsilon_\pm|\ll 1$ for large $L$. Hence we have
\be
\psi_-(\alpha_{\rm BR}^-)\simeq
\ln\left[\frac{2\epsilon_-\sinh\eta }{\cosh(3\eta)-\cosh\eta}\right]\ ,
\label{psimepsm}
\ee
which implies that if $\ln(|\epsilon_-|)={\cal O}(L)$ the last term in \fr{zeromodeEV} can contribute to ${\cal O}(L)$. In contrast, $\psi_-(\alpha_{\rm BR}^+)$ is not exponentially large in $L$ by virtue of our restriction \fr{restriction} on $\txi^+$.
In order to proceed we need to determine the large-$L$ behaviours of $\epsilon_\pm$. These can be computed from the Bethe equations \fr{BAE} for the boundary roots, which gives
\begin{align}
\ln\epsilon_\sigma&=L\epsilon^{(0)}_\sigma+\epsilon^{(1)}=
2L\ln\Big[\frac{\sinh\txi_\sigma}{\sinh(\eta+\txi_\sigma)}\Big]
-L\int_{-\frac{\pi}{2}}^{\frac{\pi}{2}}
\frac{d\alpha}{2\pi}\rho_0(\alpha)\ \psi_\sigma(\alpha)\nn
&-\int_{-\frac{\pi}{2}}^{\frac{\pi}{2}}
\frac{d\alpha}{2\pi}\rho_1(\alpha)\
\psi_\sigma(\alpha)
+\sum_{j=1}^{n_h}\psi_\sigma(\check{\lambda}_{h_n})-
\sum_{\lambda_k\in\mathcal{Z}'}\psi_\sigma(\lambda_k)\nn
&+\ln\Big[\frac{\sinh(\eta-2\txi_\sigma)}{\sinh(3\eta+2\txi_\sigma)}\Big]
+\ln\Big[\frac{\sinh(\eta+2\txi_\sigma)\sinh(\delta_++\eta)}{\sinh(\sigma\delta_-)}\Big]
+o(L^{0}),
\label{epspm}
\end{align}
where $\mathcal{Z}'_\sigma=\mathcal{Z}-\{\abr^\sigma\}$ and $\delta_\pm=\txi^+\pm\txi^-$. Substituting \fr{epspm} and \fr{psimepsm} into \fr{zeromodeEV} we then obtain
\begin{align}
\chi(\{\wta\},\abr^-)&=\ln(2)+ o(L^0)\ ,\nn
\chi(\{\wta\},\abr^-)&=-L\big[2\cosh\eta+
\int_{-\frac{\pi}{2}}^{\frac{\pi}{2}}\frac{d\alpha}{2\pi}\psi_-(\alpha)\big]+{\cal O}(L^0)\ .
\end{align}
These agree with our conjectures \fr{psip1} and \fr{psim1}. Analogous calculations can in principle be done for longer boundary strings, but the determination of their finite-size corrections becomes considerably more cumbersome. For higher values of the spin $S$ the ground state sector involves a macroscopic number of $2S$-strings. Their finite-size deviations are important as can be seen from the explicit form of the ESZM eigenvalue. This makes the calculation of the latter much more difficult than in the $S=\tfrac12$ case.

\section{Conclusions}
\label{sec:conclusions}

We have shown that the physics of strong zero mode operators for higher-spin chains is much more intricate than that of the spin-$\tfrac12$ chain studied previously. Indeed, we saw that the gapped and integrable spin-$S$ XXZ chain possesses $2S+1$ degenerate ground states for periodic boundary conditions. Thus the pairing between states in different symmetry sectors via the SZM cannot occur for integer $S$. We showed these degenerate states do not appear as a consequence of any obvious symmetry, and so naturally can be understood as a consequence of the integrable line describing a first-order phase transition between different ordered states in general. 

The integrable chains are thus highly non-generic. 
However, they have the feature of allowing an explicit construction of a candidate strong zero mode, including exact ones that commute exactly with the Hamiltonian. To this end, we generalised a transfer-matrix construction developed for the spin-$\tfrac12$ case \cite{FGVV,vernier2024strong} to these integrable spin-$S$ chains. By utilising a product of the commuting transfer matrices inherent to an integrable model, we constructed operators that are guaranteed to commute with the associated Hamiltonian. However, they differ from the spin-$\tfrac12$ case in several important ways. They are not strictly localised at the edge, and they do not square to the identity (even ignoring corrections exponentially small in the system size). Nevertheless, we showed that both properties still hold in the sense of operator norms, i.e.\ they hold for all but an exponentially small number of states relative to the full Hilbert space. Thus the ESZM with slightly weaker properties can occur even in the presence of an odd number of ground states.


Crucially, this weak form of spatial locality is still sufficient to imply the same kind of infinite coherence times previously reported for models that exhibit strong zero modes with the usual properties. We checked this assertion by doing numerics on small systems, including ones with integrability broken. The properties are very much in accord with the existence of both SZMs and ESZMs. For example, Fig~\ref{fig:autoxxz_hl} shows that infinite-temperature edge autocorrelation functions for $S$\,=\,$\tfrac12$ relax to a finite plateau at late times, indicating the presence of an ESZM. Figs~\ref{fig:autocorr} and \ref{fig:autocorr_x} show the same behaviour for the two kinds of integrable, $\mathds{Z}_2$-invariant boundary conditions in the $S$\,=\,1 chain. Tuning the boundary conditions at the ``far" edge (site $L$) in order to have equivalent boundaries turns the ESZM into a SZM. This is reflected in the behaviour of the infinite-temperature edge-autocorrelation functions shown in Figs~\ref{fig:autoxxz_hl0} (for $S$\,=\,$\tfrac12$) and \ref{fig:autocorrS1_SZM_x} (for $S$\,=\,1), which exhibit a plateau the duration of which grows with the system size $L$, before decaying at late times.
 
An open question is if strong zero modes occur in more generic models with higher spin, where the peculiar ground-state degeneracy does not occur. While an exact strong zero mode seems unlikely, there is no obstacle to having an SZM or an almost-SZM result in very long if not infinite coherence times. However, finding such operators for non-integrable models will likely need to be computer-assisted in the fashion of \cite{kemp2017long}, as a brute force approach is not practical. 

Another interesting direction to pursue is the construction of different ESZMs in integrable spin chains with $S\geq 1$. As we have seen in the $S=1$ case, the ESZM operator we constructed cannot account for the plateau values of infinite-temperature edge autocorrelation functions, which strongly suggests the existence of other conserved charges localized around the boundaries. In the $S=\tfrac32$ case we have presented some analytic evidence that points to the existence of such ESZMs. In particular, a brute-force iterative approach does yield the first two terms in the large-$\eta$ expansion of the ESZM here, in contrast to spin 1.  Moreover, the transfer-matrix approach allows for other putative ESZMs to be constructed by using higher-dimensional auxiliary spaces. While the expressions are rather unwieldy, we worked out the first two terms for $\tfrac32$ explicitly and saw that they agreed with those found in the iterative approach. We thus are optimistic that with more effort, the technical complications could be controlled. A strong motivation for doing so is that if such an SZM exists, it possibly has all the stronger properties of the spin-$\tfrac12$ mode. 

\section*{Acknowledgements}
This work was supported in part by the EPSRC under grant EP/X030881/1 (FHLE and PF). FHLE
thanks the Institut Henri Poincaré (UAR 839 CNRS-Sorbonne Université) and the LabEx CARMIN
(ANR10-LABX-59-01) for their support. 

\appendix

\section{Transfer matrices and the higher-spin Hamiltonians}

\subsection{T-system relations}
\label{app:Tsystem}

In this Appendix we describe the well-known T-system relations, which allow to construct transfer matrices $T^{(S,S')}(u)$ with arbitrary auxiliary spin from the ``fundamental'' $T(u) \equiv T^{(S,\frac12)}(u)$ \cite{ZHOU1995619,Mezincescu_1992,ZHOU1996504,Piroli_2017}. 
Throughout this Section the physical spin $S$ is fixed, and will therefore sometimes be omitted from notations. 

Introducing the functions
\begin{align}
\omega_1(u)&=\frac{\sinh(2u+2\eta)\sinh(u+\xi^+)\sinh(u+\xi^-)}{\sinh(2u+\eta)}\,, \nn
\omega_2(u)&=\frac{\sinh(2u)\sinh(u-\xi^++\eta)\sinh(u-\xi^-+\eta)}{\sinh(2u+\eta)}\,,  \nn
\phi(u) &= \left( \frac{\sinh(u+(\tfrac{1}{2}-S)\eta)}{\sinh(u+(\tfrac{1}{2}+S)\eta)} \right)^{2L}\ , \nn
f(u) &= \frac{\omega_2(u) \phi(u)}{\omega_1(u)} 
\end{align}
as well as 
\begin{equation}
\begin{aligned}
t_0(u) &= \mathds{1}  \nonumber\\
t_1(u) &= \frac{T(u)}{\omega_1(u)}  \equiv t(u) \,,  
\end{aligned}
\end{equation}
the T-system recursion relations take the form  
\be 
t_j(u) = t_{j-1}\left(u-\frac{\eta}{2}\right)t_{1}\left(u+(j-1)\frac{\eta}{2}\right)  - f\left(u + (j-3)\frac{\eta}{2}\right) t_{j-2}(u-\eta) \,, \quad j\geq 2 \,.
\label{Tsys1} 
\ee 
$t_j(u)$ corresponds (up to a normalization) to the transfer matrix with auxiliary spin $\tfrac{j-1}{2}$, namely $T^{(S,\frac{j-1}{2})}(u)$.
From (\ref{Tsys1}) one can derive the relation
\be 
t_j\left( u+\frac{\eta}{2} \right)t_j\left( u-\frac{\eta}{2} \right) 
= 
t_{j+1}( u ) t_{j-1}( u )  + \Phi_j(u) \,, \qquad j\geq 1
\ee
where 
\be 
\Phi_j(u) = \prod_{k=1}^j f\left( u - (j+2-2k)\frac{\eta}{2} \right) 
   \,.
\ee 
More generally, we have for $j \geq m$
\be  
t_j(u+m\tfrac{\eta}{2})t_j(u-m\tfrac{\eta}{2}) 
= 
t_{j+m}(u) t_{j-m}(u) + \Phi_{j-m+1}(u) t_{m-1}(u+(j+1)\tfrac{\eta}{2}) t_{m-1}(u-(j+1)\tfrac{\eta}{2})  \,.
\ee

\subsection{Explicit expression for the integrable \sfix{spin-$S$} Hamiltonian}
\label{app:suq2spins}
The Hamiltonians are parametrized by an anisotropy parameter $\eta$ and are conveniently expressed in terms of quantum group generators $S_j^{z,\pm}$, acting locally on the spin-$S$ on site $j$ as  
\be 
\begin{split}
S^z_j |m \rangle_j &= m |m \rangle_j \,, \qquad m = -S,\ldots,S 
\\
S^\pm_j |m \rangle_j &=  \sqrt{ [S+1\pm m] [S \mp m]} | m \pm 1 \rangle_j \, ,
\end{split}
\label{eq:Uqsl2}
\ee
where we have introduced notations
\be
[x] := \frac{\sinh (\eta x)}{\sinh \eta}= \frac{q^x-q^{-x}}{q-q^{-1}}\ ,\quad
q=e^\eta\ .
\ee
The spin-$S$ Hamiltonian then reads
\be 
H^{(S)} = h_1^{(S)} +  \sum_{j=1}^{L-1} h_{j,j+1}^{(S)} 
+ h_L^{(S)}+{\rm const} \,,
\label{H_S}
\ee 
where the local densities are given by \cite{bytsko2003integrable} 
\begin{align}
h_{j,j+1}^{(S)}  &=  4(\cosh\eta)^{2S-1}\sinh\eta \sum_{j=1}^{2S} \left[ 
\left(
\sum_{k=1}^j \frac{\cosh(\eta k)}{\sinh(\eta k)}
 \right)
 \prod_{l=0, \neq j}^{2S} \frac{2 X^{(S)}_{j,j+1}  - [l][l+1]}{[j-l][j+l+1]}
\right] \,, 
\label{hSgeneric}\\
X^{(S)}_{j,j+1} &= q^{S_j^z}\Big[
\frac{1}{2} S_j^+ S_{j+1}^- + \frac{1}{2} S_j^- S_{j+1}^+ + \frac{\sinh(\eta S_j^z)}{\sinh \eta}\frac{\sinh(\eta S_{j+1}^z)}{\sinh \eta} \cosh(S\eta) \cosh((S+1)\eta) \nn
&\qquad\quad+ \cosh(\eta S_j^z) \cosh(\eta S_{j+1}^z) \frac{\sinh(\eta S)}{\sinh \eta}\frac{\sinh((S+1)\eta)}{\sinh \eta}
\Big] q^{-S_{j+1}^z} \ .
\end{align}

The boundary terms $h_1^{(S)}$, $h_L^{(S)}$ act diagonally on the first and last spin respectively, with coefficients depending on a pair of boundary parameters $\xi^\pm$ that are fixed by requiring commutation with a family of transfer matrices (see below). Up to an overall constant the boundary terms follow from
\begin{align}
\langle m|h_1^{(S)}|m\rangle -\langle m-1|h_1^{(S)}|m-1\rangle  &= 
4(\cosh\eta)^{2S-1}
\frac{(1-q) (q+1)}{2 \left(q-e^{2 \xi^+} q^{2
   m}\right)}\ ,\quad  -S<m\leq S\ ,\nn
\langle m|h_L^{(S)}|m\rangle -\langle m-1|h_L^{(S)}|m-1\rangle  &= 
4(\cosh\eta)^{2S-1}
 \frac{(1-q) (q+1)}{2  \left(e^{-2 \xi^-} q^{-2
   (m-1)}-q\right)} \ .
   \label{eq:boundaryterms}
\end{align}
The asymmetry between the expressions for the left and right boundary terms can be related to the asymmetry of the bulk densities \eqref{hSgeneric} under the exchange $j\leftrightarrow j+1$. This arises because \eqref{hSgeneric} contains contributions that cancel in the bulk when summed over $j$, and only give contributions on the two boundaries.
All asymmetries disappear in the resulting global Hamiltonian \eqref{H_S}, which is invariant under space reflection and exchange of the boundary parameters $\xi^+ \leftrightarrow \xi^-$. This can be seen in the explicit expressions for $S=\tfrac12$ and $S=1$ given in \fr{Hopenper} and \fr{boundaryfieldshalf},\fr{bulkhamiltonianspin1}.

\subsection{Explicit expression for \sfix{$T^{(\frac12,1)}(\mu)$}}
\label{app:halfone}
In this Appendix we present explicit expressions for the L-operator and K-matrices used to construct the transfer matrix
$T^{(\frac12,1)}(\mu)$. The L-operator is
\be
 {\cal L}(\mu)=
\begin{pmatrix}
  1 & 0 & 0 & 0 & 0 & 0\\
  0 &c(\mu) &
  a(\mu)
  &0&0&0\\
  0&a(\mu) &
  b(\mu)&0&0&0\\
  0&0&0&b(\mu)&
  a(\mu) &0\\
  0&0&0&a(\mu)&
  c(\mu)  &0\\
  0&0&0&0&0&1
\end{pmatrix}\ ,
\ee
where we have defined
\be
a(\mu)=\sqrt{2\cosh\eta}\frac{\sinh\eta}{\sinh(\frac{3\eta}{2}+\mu)}\ ,\quad
b(\mu)=\frac{\sinh(\frac{\eta}{2}+\mu)}{\sinh(\frac{3\eta}{2}+\mu)}\ ,\quad
c(\mu)=-\frac{\sinh(\frac{\eta}{2}-\mu)}{\sinh(\frac{3\eta}{2}+\mu)}.
\ee
The corresponding diagonal K-matrices are
\begin{align}
\big(K^-(\mu)\big)_{11}&=
\sinh(\mu-\frac{\eta}{2}+\xi^-)\sinh(\mu+\frac{\eta}{2}+\xi^-)\ ,\nn
\big(K^-(\mu)\big)_{22}&=
-\sinh(\mu-\frac{\eta}{2}-\xi^-)\sinh(\mu-\frac{\eta}{2}+\xi^-)\ ,\nn
\big(K^-(\mu)\big)_{33}&=\sinh(\mu-\frac{\eta}{2}-\xi^-)\sinh(\mu+\frac{\eta}{2}-\xi^-)\ ,\nn
\big(K^+(\mu)\big)_{11}&=
\sinh(\mu+\frac{\eta}{2}+\xi^+)\sinh(\mu+\frac{3\eta}{2}+\xi^+)\ ,\nn
\big(K^+(\mu)\big)_{22}&=
-\sinh(\mu+\frac{3\eta}{2}+\xi^+)\sinh(\mu+\frac{3\eta}{2}-\xi^+)\ ,\nn
\big(K^+(\mu)\big)_{33}&=\sinh(\mu+\frac{\eta}{2}-\xi^+)\sinh(\mu+\frac{3\eta}{2}-\xi^+)\ .
\end{align}

\section{No ESZM operators from transfer matrices \sfix{$T^{(\frac{1}{2},1)}(u)$}}
\label{app:ShalfSprime}

The L-operator and K-matrices for $S=\frac{1}{2}$ and $S'=1$ are given in Appendix \ref{app:halfone}. From these we construct 
the family of transfer matrices
\be
T^{(\frac{1}{2},1)}(\mu)={\rm Tr}\left[K^+(\mu){\cal L}_1(\mu)\dots{\cal L}_L(\mu)K^-(\mu){\cal L}_L(\mu)\dots{\cal
  L}_1(\mu)\right].
\ee
In order to assess whether $T^{(\frac{1}{2},1)}(\mu)$ can be used to obtain an ESZM operator we proceed as follows:
\begin{enumerate}
\item{} We fix $\xi^+$ to the required value to obtain an ESZM operator for $T^{(\frac{1}{2},\frac{1}{2})}(\mu)$
  \be
\xi^+=\frac{i\pi}{2}\ .
  \ee
This ensures that the charge we construct is compatible with the known ESZM.
\item{} We require that there exists a spectral parameter $u^*$ such that
  \be
  {\cal L}(u^*)K^-(u^*){\cal L}(u^*)={\rm const }\ K^-(u^*)\ .
\label{requirement}
\ee
The motivation for this requirement is that it implies that the derivative of the transfer matrix can be written as a sum over operators $\widetilde{\Psi}_j$ that act non-trivially only on sites $1$ to $j$, and which are obtained in the same way as in section \ref{ssec:TShalf}
\be
{T^{(\frac{1}{2},1)}}'(u^*)\propto\sum_{j=1}^L\widetilde{\Psi}_j(\eta)+{\rm const}\ .
\ee

We find that there are only two solutions for $u^*=u^*_{1,2}$ to \fr{requirement}
\be
u^*_1=\frac{i\pi}{2}\ ,\quad u^*_2=0.
\ee
For these we have
\be
T^{(\frac{1}{2},1)}(u^*_{1,2})\propto\mathds{1}\ .
\ee
\item{}
We examine the locality properties of $\widetilde{\Psi}_j(\eta)$ in the strong coupling
limit $\eta\to\infty$. For $u^*_1$ we then find, after
dividing by an overall factor $2[\sinh(\frac{3\eta}{2})\cosh(\frac{\eta}{2})]^2$
\begin{align}
\widetilde{\Psi}_1(\infty)&=\mathds{1}\ ,\
\widetilde{\Psi}_2(\infty)=\mathds{1}-e_1^{11}e_2^{11}-e_1^{22}e_2^{22}\ ,\
\widetilde{\Psi}_3(\infty)=\mathds{1}-e_1^{11}e_2^{11}e_3^{11}-e_1^{22}e_2^{22}e_3^{22}\ ,\nn
\widetilde{\Psi}_4(\infty)&=\mathds{1}-e_1^{11}e_2^{11}e_3^{11}e_4^{11}-e_1^{22}e_2^{22}e_3^{22}e_4^{22}
-e_1^{11}e_2^{22}e_3^{11}e_4^{11}\nn
&\qquad -e_1^{11}e_2^{22}e_3^{22}e_4^{22}
-e_1^{22}e_2^{11}e_3^{22}e_4^{22}-e_1^{22}e_2^{11}e_3^{11}e_4^{11}\ .
\end{align}
These do not have the desired spatial locality properties as
\begin{align}
&\frac{1}{2^2}{\rm Tr}\big(\widetilde{\Psi}^2_2(\infty)\big)=\frac{2}{2^2}\ ,\
\frac{1}{2^3}{\rm Tr}\big(\widetilde{\Psi}^2_3(\infty)\big)=\frac{6}{2^3}\ ,\
\frac{1}{2^4}{\rm Tr}\big(\widetilde{\Psi}^2_4(\infty)\big)=\frac{10}{2^4}\ ,\nn
&\frac{1}{2^5}{\rm Tr}\big(\widetilde{\Psi}^2_5(\infty)\big)=\frac{22}{2^5}\ ,\
\frac{1}{2^6}{\rm Tr}\big(\widetilde{\Psi}^2_6(\infty)\big)=\frac{42}{2^6}\ .
\end{align}
The coefficients appear to follow the sequence
\be
a_n=a_{n-1}+2a_{n-2}\ ,\quad a_0=0\ ,\ a_1=2\ ,
\ee
which would imply that the Hilbert-Schmidt norm of $\widetilde{\Psi}_j$ approaches a constant for large $j$ rather than decaying exponentially in $j$. We conclude that the conserved charge constructed in this way does not constitute an ESZM operator because it lacks the required spatial locality properties.
The same holds for the strong-coupling expansion for $u^*_2$.
\end{enumerate}
\section{Sums over real Bethe roots in the \sfix{$S=\frac{1}{2}$} XXZ chain}
\label{app:sumtoint}
In this appendix we summarize some standard formulas for turning sums over real Bethe roots in the $S=\frac{1}{2}$ XXZ chain with $\Delta>1$ into integrals in the limit of large system sizes $L$.
We consider a solution of the form set out in the main text, i.e. containing
$M_r$ real roots $\alpha_j$ (where $M_r$ scales with $L$), $M_c$ roots ${\cal Z}=\{\lambda_j\}$ with non-zero imaginary part, where $M_c$ is fixed when $L$ increases. 
The main identity is (\emph{cf} (B.6) of \cite{Grijalva2019open})
\begin{align}
\sum_{k=1}^{M_r}f(\alpha_k)&=L
\int_{-\frac{\pi}{2}}^{\frac{\pi}{2}}\frac{d\alpha}{2\pi}
f(\alpha) \rho_0(\alpha)\nn
&+\int_{-\frac{\pi}{2}}^{\frac{\pi}{2}}\frac{d\alpha}{2\pi}
f(\alpha) \rho_1(\alpha)
-\frac{f(0)+f(\pi/2)}{2}-\sum_{n=1}^{n_h}f(\check{\lambda}_{h_n})+{\cal O}(L^{-\infty})\ .
\label{sumtoint}
\end{align}
Here $\{\check{\lambda}_{h_n}\}$ a set of holes as defined in \fr{holes} and $\rho_a(\alpha)$ are solutions to the linear integral equations
\begin{align}
\rho_a(\alpha)&=\rho_a^{(0)}(\alpha)-\int_{-\frac{\pi}{2}}^{\frac{\pi}{2}}\frac{d\mu}{2\pi}
  \phi'(\alpha-\mu,\eta)\ 
  \rho_a(\mu)\ .
\label{rho01}
\end{align}  
Here the driving terms are given by
\begin{align}
\rho_0^{(0)}(\alpha)=& \phi'(\alpha,\frac{\eta}{2})\ ,\nn
\rho_1^{(0)}(\alpha)=&
\frac{1}{2}\Big[-\sum_{\sigma=\pm}\phi'(\alpha,\frac{\eta}{2}+\xi_\sigma)+2\phi'(2\alpha,\eta)\nn
&+\phi'(\alpha,\eta)+\phi'(\alpha+\pi/2,\eta)-\sum_{\lambda_k\in\mathcal{Z}}\Phi'(\alpha,\lambda_k)+\sum_{j=1}^{n_h}\Phi'(\alpha,\check{\lambda}_{h_n})\Big],
\label{rho}
\end{align}
where we have defined
\begin{align}
\Phi(\alpha,\lambda)=\phi(\alpha-\lambda,\eta)+\phi(\alpha+\lambda,\eta)\ ,\quad
\phi(\alpha,\eta)=i\ln\left[-\frac{\sin(\alpha+i\eta)}{\sin(\alpha-i\eta)}\right].
\end{align}
The integral equation for $\rho_0(\alpha)$ is easily solved by Foruier methods
\begin{align}
\rho_0(\alpha)&=\sum_{k\in\mathbb{Z}}\frac{e^{2ik\alpha}}{\cosh(k\eta)}\ .
\end{align}

\section{Numerical results for \sfix{$S=\tfrac12$} ESZM eigenvalues on Bethe states}
\label{app:numerics}
In this appendix we report numerical results on short chains for solving the Bethe equations for $S=\tfrac12$ and evaluating the corresponding ESZM eigenvalue. In Table~\ref{tab:L8N4} we present results for $L=2N=8$ and $\Delta=4$, $h_L=8$, $h_1=0$. The solutions of the Bethe equations can be understood in terms of an appropriate string hypothesis \cite{takahashi_1999,babelon1983analysis,kapustin1996surface,mossel2010relaxation,Grijalva2019open}. In order to identify the "string content" of a given solution we introduce the following notations:
\begin{itemize}
\item{} $\rm r$ denotes a real root of the Bethe equations.
\item{} $\rm 2s$ denotes a 2-string, i.e. a pair of roots
\be
\alpha_{1,2}=\alpha\pm \frac{i\eta}{2}+\delta^\pm\ ,\quad \alpha\in\mathbb{R}\ ,
\ee
where $\delta^\pm$ are small deviations from an ideal string.
\item{} $\rm b^\pm$ denote boundary roots associated with the right/left boundary, i.e. roots approximately given by \fr{boundarystrings} with $n=1$.
\item{} $\rm b2s^\pm$ denote boundary 2-strings associated with the right/left boundary, i.e. roots approximately given by \fr{boundarystrings} with $n=2$.
\end{itemize}
\begin{table}[ht]
\begin{tabular}{|c|c|c|c|c|c|c|c|}
\hline
E&$\alpha_1$&$\alpha_2$&$\alpha_3$&$\alpha_4$& $\Psi$ & roots \\
\hline
-65.88&0.5034i&0.5814&0.2741&0.9872&0.9990&b$^+$,r,r,r\\
\hline
-59.84&1.571+1.031 i&0.5034i&0.2694&2.571&-0.9989&b$^-$,b$^+$,r,r\\
\hline
-56.89&1.571+1.031 i&0.5034 i& 0.2682&2.175&-0.9989 &b$^-$,b$^+$,r,r\\
\hline
-54.57&1.571+1.031i&0.5034i&0.5670&2.177&	-0.9990&b$^-$,b$^+$,r,r\\
\hline
-52.90&0.5034i&2.562i& 0.2160&0.4523&	0.9999&b2s$^+$,r,r\\
\hline
-50.85&0.5034i&1.049-1.032i&$\alpha_2^*$&0.2677&	0.9989 &b$^+$,2s,r\\
\hline
-49.22&1.571+1.031i&0.2665&0.5649&2.180&-0.9989 &b$^-$,r,r,r\\
\hline
-48.91&0.5034i&2.565i& 0.4508&2.408&0.9999&b2s$^+$,r,r\\
\hline
-41.75&1.571+1.032i&1.571+3.093i&0.5034i&0.480085&	-1.0&b2s$^-$,b$^+$,r\\
\hline
-38.77&1.571-1.032i&0.5034i&2.566i&0.7188&-0.9998& b$^-$,b2s$^+$,r\\
\hline
-31.74&1.571+1.032i&1.571+3.092i&0.2301&2.017&-1.0&b2s$^-$,r,r\\
\hline
-28.92&1.571+1.032i&0.6163+1.032i&$\alpha_2^*$&0.9689&-0.9989&b$^-$,2s,r\\
\hline
-27.91&1.571+1.032i&1.571+3.091i&0.7602&1.122&-1.0&b2s$^-$,r,r\\
\hline
-22.88&0.5169+1.032i&1.185+1.032i&$\alpha_1^*$&$\alpha_2^*$&0.9989&2s,2s\\
\hline
\end{tabular}
\caption{\small Solutions to the BAE for $\Delta=4$, $h_L=8$, $L=2N=8$.}
\label{tab:L8N4}
\end{table}
The numerical results are in agreement with our claim that the eigenvalue of the SZM operator is $-1$, if the solution contains a boundary string associated with the right boundary $\txi^-$, and $+1$ otherwise.

\bibliography{bibliography}

\end{document}